\documentclass[a4paper,11pt]{article}
\pdfoutput=1 

\usepackage{jheppub} 

\usepackage[T1]{fontenc} 

\newcommand{\ga}{\gamma}
\newcommand{\be}{\begin{eqnarray}}\newcommand{\ed}{\end{eqnarray}}
\newcommand{\non}{\nonumber}
\newcommand{\ra}{\rangle}\newcommand{\la}{\langle}

\title{\boldmath Photon Polarization in the $b\to s\ga$ processes \\in the Left-Right Symmetric Model}

\author[a,b,c]{Fu-Sheng Yu,}
\author[b]{Emi Kou,}
\author[c]{Cai-Dian L\"u}

\affiliation[a]{School of Nuclear Science and Technology, Lanzhou University, Lanzhou 730000, People's Republic of China}
\affiliation[b]{Laboratoire de
l'Acc$\acute{\rm e}$l$\acute{\rm e}$rateur Lin$\acute{\rm e}$aire,
    Universit$\acute{\rm e}$ Paris-Sud 11, CNRS/IN2P3(UMR 8607) 91405
    Orsay, France}
\affiliation[c]{Institute of High Energy Physics and Theoretical Physics
Center for Science  Facilities, Chinese Academy of Sciences, Beijing
100049, People's Republic of China}

\emailAdd{yufsh@lzu.edu.cn}
\emailAdd{kou@lal.in2p3.fr}
\emailAdd{lucd@ihep.ac.cn}

\abstract{The circular-polarization of the photon in the radiative $B$ decays
is dominantly left-handed in the Standard Model (SM), but the right-handed polarization may be enhanced in some new physics models with right-handed currents, such as the Left-Right Symmetric Model (LRSM). In this article, we investigate how large this {\it wrong polarization} could be within the allowed parameter space of the LRSM. We show that in the LRSM, the right-handed polarization of the photon in the $b\to s\ga$ processes could be largely enhanced by the  $W_L-W_R$ mixing contributions because of the helicity flip on the internal top quark line of the penguin diagrams and the enhancement by the CKM factor $V_{ts}^R/V_{ts}^L$.
We  discuss the sensitivity of the proposed methods to determine  the $b\to s\gamma $ photon polarization to the LRSM as well as  their complementary role  compared to the direct search of right-handed charged gauge bosons at LHC.
}


\begin{document}
\maketitle
\flushbottom

\section{Introduction}
The circular-polarization of the photon in the radiative $B$ decays
has an unique sensitivity to the physics beyond the Standard Model (SM).
There have been  various methods proposed to measure this polarization but its precise measurement has never been achieved so far. Therefore,
it is an important challenge for the on-going experiments such as LHCb or the future projects such as Belle II experiment. For this reason, there are many theoretical works in this subject recently~\cite{theoryworks1,theoryworks2,theoryworks3,theoryworks4,theoryworks5,theoryworks6,theoryworks7,theoryworks8,theoryworks9,theoryworks10}.

The $b \to s \gamma$ processes come from the magnetic-operators, $\overline{s}_{L}\sigma_{\mu\nu}b_{R}$ or  $\overline{s}_{R}\sigma_{\mu\nu}b_{L}$. The photons induced by these operators have left- and right-handed polarization, respectively.
In the SM, these operators are induced by the loop diagrams with the $W$ bosons.
Due to the fact that  the $W$  bosons couple only left-handedly to the quarks, the first operator receives a bottom quark mass factor, {\it i.e}. $m_b\overline{s}_{L}\sigma_{\mu\nu}b_{R}$, while the second one has a strange quark mass factor, {\it i.e}. $m_s\overline{s}_{R}\sigma_{\mu\nu}b_{L}$.
As a result, the circular-polarization of the photon in the $b\to s\gamma$ processes is predicted to be predominantly left-handed in the SM.
Thus, a measurement of the photon polarization can serve as a test of a fundamental property of the SM. On the other hand, an observation of {\it wrong polarization} (right-handed in $\bar{b}$ quark decay or left-handed in ${b}$) is a clear indication from the physics beyond the SM, in particular, from the models which contain right-handed couplings (see \cite{SMuncertainties1,SMuncertainties2,SMuncertainties3,SMuncertainties4} for the SM uncertainty).

In this article, we investigate one of the models in this category, the Left-Right Symmetric Model (LRSM).
This model contains a $W$-type boson which has right-handed charged coupling to the quarks.
The LHC experiments, namely ATLAS and CMS, are intensively searching for this type of particles and the latest limit on its mass and coupling is obtained as: $g_R/g_L m_{W_2}>2.5$\ TeV~\cite{newlhcconstraints1,newlhcconstraints2}.
The right-handed charged couplings to different quarks, which are represented by the right-handed CKM matrix elements, can be  constrained, on the other hand, by various flavour physics experiments.
By now, the minimal LRSM with a symmetry between the left- and right-handed sector \cite{Maiezza:2010ic}, {\it i.e.} the so-called manifest- and pseudo-manifest-models  motivated by the spontaneous CP violation \cite{manifest1,manifest2,pseudomanifest1,pseudomanifest2,pseudomanifest3,pseudomanifest4}, are becoming difficult to realize~\cite{Zhang:2007da,Ball:1999mb,Barenboim:2001vu}, whereas in this work,  we use a more general right-handed CKM matrix.
We first revisit the constraints on these parameters of the LRSM, namely, new particle masses, their couplings and the right-handed CKM matrix elements, from various flavour experiments ($K$, $B_d$, $B_s$ oscillation and $B\to X_s\gamma$ branching ratio). Then, we investigate how large the {\it wrong polarization} could be using the allowed ranges of these parameters. We show the effect of the so-called chiral-enhancement factor, which is expected to play a crucial role in enhancing the  {\it wrong polarization}:
the two $W$ boson mass eigenstates in this model being mixing states of left- and right-handed $W$ bosons can induce the chirality change without a flip on the external $b$ or $s$ quark lines, which provide an internal quark mass factor $m_t$ instead of external quark mass factor mentioned above. As a result, the operator  $\overline{s}_{R}\sigma_{\mu\nu}b_{L}$ is enhanced by a factor $m_t/m_s$ with respect to the SM.

At the end, we discuss the prospect of the LHCb and the future Belle II experiments for constraining the LRSM parameters by using some of their observables,  the time dependent CP asymmetry of $B\to K_S\pi^0\gamma$~\cite{kpiga1,kpiga2}, the angular distribution of $B\to K^*l^+l^-$ (at low $q^2$)~\cite{kstarll1,kstarll2}, and $B\to K_1\gamma$~\cite{k1gamma1,k1gamma2,k1gamma3,k1gamma4}.

In section 2, we introduce the LRSM where we also give our model for the right-handed CKM matrix.
In section 3 and 4, we describe the $b\to s\gamma$ and meson mixings in the LRSM. We show our numerical results in section 5 and conclude in section 6.

\section{Left-Right Symmetric Model}

The Left-Right Symmetric Model (LRSM) is based on the extended gauge group
$SU(2)_L\times SU(2)_R\times U(1)_{\tilde Y}$ which involves additional charged and  neutral gauge
bosons \cite{patisalam1,patisalam2,patisalam3,patisalam4}.
The electric charge can be also extended as $Q=T_{L3}+T_{R3}+\tilde{Y}$. Then, for the ordinary quarks and leptons,  the hypercharge gets a physical meaning, i.e. $\tilde{Y}=(B-L)/2$ in this model, where $B$ and $L$ are baryon and lepton numbers.

The left-handed fermions are  $SU(2)_L$ doublets and $SU(2)_R$ singlets as in the SM while the right-handed fermions are  $SU(2)_R$ doublets and $SU(2)_L$ singlets. Thus, the charge assignments $(T_{L3}, T_{R3}, \tilde{Y})$ of left- and right-handed quarks and leptons yields:
 \be
 Q_L\equiv\left(\begin{array}{cc} u_L \\ d_L \end{array}\right)\sim\left({1\over2},0,{1\over6}\right)&,&
 ~~~~~~Q_R\equiv\left(\begin{array}{cc} u_R \\ d_R
 \end{array}\right)\sim\left(0,{1\over2},{1\over6}\right),\\
  L_L\equiv\left(\begin{array}{cc} \nu_L \\ \ell_L \end{array}\right)\sim\left({1\over2},0,{-{1\over 2}}\right)&,&
 ~~~~~~L_R\equiv\left(\begin{array}{cc} \nu_R \\ \ell_R
 \end{array}\right)\sim\left(0,{1\over2},{-{1\over 2}}\right),
 \ed

 The gauge symmetry in the LRSM is spontaneously broken in two steps
\be\label{SB}
SU(2)_L\times SU(2)_R \times U(1)_{\tilde{Y}} \rightarrow SU(2)_L\times U(1)_Y \rightarrow U(1)_{\rm EM}.
\ed
The first step $SU(2)_R\times U(1)_{\tilde{Y}}\rightarrow U(1)_Y$ is parity and $B-L$ violating while the second step is equivalent to the standard electroweak symmetry breaking.
Let us first see the scalar multiplet $\Phi$, which triggers the second step symmetry breaking.
Consulting the Yukawa interaction of the form,  $\overline{Q}_L\Phi Q_R$, $\Phi$ should be a $2\times 2$ unitary matrix.
Moreover,  to be invariant under $SU(2)$ transformation, this term requires $\Phi $ to be  bi-doublet scalar fields with charge assignment:
 \be
 \Phi\equiv\left(\begin{array}{cc} \varphi_1^0 & \varphi_2^+ \\
                              \varphi_1^- & \varphi_2^0  \end{array}\right)\sim\left({1\over2},{1\over2},0\right).
 \ed
The scalar field $\Phi$ can not trigger the first step symmetry breaking because (i), $\Phi$
couples to both of $SU(2)_L$ and $SU(2)_R$ and does not distinguish
these two groups. Therefore $\Phi$ could not break parity.
 (ii) $\Phi$ does not couple to $U(1)_{\tilde{Y}}$ which would be unbroken and would leave a massless gauge boson which is not observed.
 Thus, we must introduce other scalar multiplets to break the parity symmetry, namely the $SU(2)_R$, and also $U(1)_{\tilde{Y}}$.
 In particular, the scalar multiplet with charge $B-L=2$ is attractive since it can generate right-handed Majorana neutrino masses \cite{neutrinomass1,neutrinomass2,neutrinomass3}. As a result, we introduce the scalar triplet:
\be
\Delta_R\equiv\left(\begin{array}{cc} \delta_R^+/\sqrt{2} & \delta_R^{++} \\ \delta_R^0
        &-\delta_R^+/\sqrt{2}\end{array}\right)\sim\left(0,1,2\right),\non
\ed
Note that in this model another scalar field $\Delta_L\sim(1,0,2)$ is also introduced in order to ensure the parity conservation at a high energy before the symmetry is broken.

Now the symmetry breaking~(\ref{SB}) can be undertaken by the non-zero vacuum expectation values (VEV). It is known that two of four complex phases can be eliminated and remaining two phases are assigned conventionally to the VEV of $\Phi$, $\Delta_L$ and $\Delta_R$ \cite{Zhang:2007da}:
 \be
 \la\Phi\ra=\left(\begin{array}{cc} \kappa & 0 \\ 0 & \kappa'e^{i\omega}
                    \end{array}\right),\label{vev:phi}
~~~~
 \la\Delta_{L}\ra=\left(\begin{array}{cc} 0 & 0 \\ v_{L}e^{i\theta_L} & 0
                    \end{array}\right),
                    ~~~~
 \la\Delta_{R}\ra=\left(\begin{array}{cc} 0 & 0 \\ v_{R} & 0
                    \end{array}\right),
                     \ed
where, $v_{L,R}, \kappa$ and $\kappa'$ are real numbers.
 The symmetry breaking of $SU(2)_R \times U(1)_{\tilde{Y}}\rightarrow U(1)_Y$ is
  achieved at a high scale, i.e. multi-TeV, by non-zero $\la \Delta_R\ra$. If $v_L\neq v_R$, the left-right symmetry is spontaneously broken.
The standard electroweak symmetry breaking is triggered by non-zero $\kappa, \kappa'$.

There are some hierarchies among the vacuum expectation values
$\kappa$, $\kappa'$ and $v_{L,R}$. First of all,
$\kappa,\kappa', v_L \ll v_R$ is needed to suppress the right-handed
currents at low energy scales.
On the other hand, we would expect another hierarchy $v_L\ll v_R $
in order to generate the neutrino masses through the see-saw mechanism, namely  $v_L\lesssim$ a few MeV \cite{neutrinomass1,neutrinomass2,neutrinomass3} for $v_R\sim$ multi-TeV.
And also the constraints from electroweak $\rho$-parameter requires $v_L\lesssim10$GeV\cite{Amaldi:1987fu}.
In this work, we shall use the limit $v_L\to0$, which is usually used in literatures.
Note that the Higgs potential allows such a limit since  $v_L\propto \kappa^2/v_R$ \cite{neutrinomass3}.
Therefore the phase $\theta_L$ has no physical consequence, while $\omega$ could trigger a spontaneous CP violation.
According to the hierarchy described above, we introduce an expansion parameter $\epsilon$ as:
\be\label{eq:v} \epsilon=v/v_R, \quad {\rm with}\quad v^2=\kappa^2+\kappa'^2, \ed
where $v=174$ GeV is the scale of the standard electroweak symmetry breaking and $v_R=\mathcal{O}(\text{TeV})$ as discussed above.
The ratio of $\kappa$ and $\kappa'$ is defined by the usual parameter $\beta$, i.e.
\be\label{eq:tanbeta}
\kappa=v \sin\beta,~~~\kappa'=v \cos\beta,~~~\tan\beta={\kappa\over\kappa'}.
\ed
Then $\tan\beta$ is a free parameter in this model.
 $\tan\beta\neq1$ is required by the difference of the masses of the fermions.
On the other hand, the mass hierarchy $m_t\gg m_b$ implies large $\tan\beta$.
However, the large value of $\tan\beta$  as $\mathcal{O}({m_t\over m_b})$ in some literatures,  is disfavored by the electroweak precision observables \cite{Blanke:2011ry}. $1<\tan\beta<10$ is used in \cite{Blanke:2011ry}.
 In this work, we consider the range of $\tan\beta >1$.

In the low energy processes, the discrete left-right symmetry breaks down, then the gauge coupling constants $g_L$ and $g_R$ are in general unequal, $ g_L\neq g_R$. As $g_L$ is the coupling constant in the SM, the ratio of $r\equiv g_R/g_L$ is not allowed to be arbitrarily large, otherwise the interactions between right-handed gauge bosons and fermions would become nonperturbative.
Having the latest direct search result, $r~ m_{W_2}>2.5$ TeV~\cite{newlhcconstraints1,newlhcconstraints2}, and  assuming $v_R$ to be muti-TeV or higher, we use $r$ less than $~2$ in this work.

The charged gauge bosons are mixture of the mass eigenstates,
 \be\label{eq:Wmxingm}
 \left(\begin{array}{cc} W_L^- \\ W_R^-  \end{array}\right)
 = \left(\begin{array}{cc} \cos\zeta & -\sin\zeta e^{iw} \\ \sin\zeta e^{-iw} & \cos\zeta \end{array}\right)
  \left(\begin{array}{cc} W_1^-  \\ W_2^-  \end{array}\right),
  \ed
where the masses of $W_1^\pm$ and $W_2^\pm$ up to the order of $\mathcal{O}(\epsilon^2)$ are
\be
M_{W_1}\approx{g_L v\over\sqrt 2}(1-\epsilon^2\sin^2\beta\cos^2\beta),~~
M_{W_2}\approx g_R v_R(1+{1\over4}\epsilon^2),
\ed
and the mixing angle
\be\label{eq:sinzeta}
\sin\zeta\approx{g_L\over g_R}{|\kappa||\kappa'|\over
 v_R^2}={g_L\over g_R}{1\over2}\epsilon^2\sin2\beta\approx{M_{W_1}^2\over M_{W_2}^2} {g_R\over g_L}\sin2\beta.
\ed
The full Lagrangian for the neutral currents and the charged currents associated with gauge bosons and Goldstone bosons in the LRSM are given in Appendix \ref{app:Lagrangian}.

In the LRSM, there are charged Higgs bosons $H^\pm$ and heavy flavor-changing neutral Higgs bosons $H^0$ and $A^0$.
In \cite{Zhang:2007da}, it is shown that the masses of these bosons are nearly the same.
To the leading order, their masses are equal to each other \cite{Blanke:2011ry},
\be
M_{H^\pm}=M_{H^0}=M_{A^0}.
\ed
For simplicity, we use $M_H$ representing the masses of charged and heavy neutral Higgs bosons.
The Lagrangians for the interactions between $H^0$, $A^0$, $H^{\pm}$ and fermions are given in Appendix \ref{app:Lagrangian}.
As we will see later-on, the tree-level flavor changing neutral current due to $H^0$ and $A^0$ will affect the $\Delta F=2$ processes very much unless $M_H$ is sufficiently large~\cite{Blanke:2011ry}. In this work, we consider the cases of $M_H=20$TeV and $50$TeV.  For such heavy mass, the contributions on $b\to s\ga$ and $\Delta F=2$ processes from charged Higgs in the loop diagrams become negligibly small.

Concerning the quark mixing matrices (CKM matrices), we have one left-handed coupling $V_{\rm CKM}^L$ and one right-handed $V_{\rm CKM}^R$ for the left- and right-handed quarks respectively.
We take $V_{\rm CKM}^L$ the same as the Standard Model one, defined by usual three rotation angles and one phase.
In this way, the right-handed CKM matrix $V_{CKM}^R$ is in general written by  nine parameters remained after imposing the unitarity condition. There are many models on the right-handed CKM matrix.
A general study of the quark sector is done in the minimal Left-Right Symmetric Model, where there is a symmetry between the left- and right-handed sector \cite{Maiezza:2010ic}.
The authors in \cite{Maiezza:2010ic} shows that the only possibilities in the minimal LRSM are  either P or C (LR+parity or LR+charge conjugation), with the consequences that $V_{\rm CKM}^R$ is equal or very close to $V_{\rm CKM}^L$. These two cases correspond to the scenarios in the early works of LRSM with the ideas of spontaneous CP violation.
In the so-called manifest LRSM\cite{manifest1,manifest2}, where the VEVs in the Higgs sector are all real, the right-handed matrix is exactly the same as the left-handed one, $V_{CKM}^R=V_{CKM}^L$, and hence there are only explicit CP violations from the phases in CKM matrices.
While in the pseudo-manifest LRSM\cite{pseudomanifest1,pseudomanifest2,pseudomanifest3,pseudomanifest4}, where the Yukawa couplings are taken to be real, the right-handed CKM matrix is related to the left-handed one by diagonal phase matrices $K^{u,d}$, $V_{CKM}^R=K^uV_{CKM}^L K^{d^\dag}$, leading to spontaneous CP violation from the complex Higgs VEV.
Both scenarios confront strong constraints from the the flavor and Higgs physics \cite{Zhang:2007da,Beall:1981ze,Ball:1999mb,Barenboim:2001vu}.
More general right-handed CKM matrix have been studies in \cite{Maiezza:2010ic,Zhang:2007da,Blanke:2011ry,generalVR1,generalVR2,generalVR3,Langacker:1989xa}.
Motivated by the $K^0-\overline K^0$ mass difference, Langacker and Sankar proposed two simple formulae of right-handed CKM matrix \cite{Langacker:1989xa},
\be
V_{(A)}^R=\left(\begin{array}{ccc}1&0&0\\0&c_\alpha&\pm s_\alpha\\0&s_\alpha&\mp c_\alpha\end{array}\right),~~~~
V_{(B)}^R=\left(\begin{array}{ccc}0&1&0\\c_\alpha&0&\pm s_\alpha\\s_\alpha&0&\mp c_\alpha\end{array}\right),
\ed
where $c_\alpha\equiv\cos\alpha$, $s_\alpha\equiv\sin\alpha$ with $\alpha$ an arbitrary angle. As we will see in the following sections, for the purpose of this work, i.e. enhancing the  right-handed polarization of the photon in $b\to s\ga$, the scenario (A) is more interesting as it keeps a possibility to have a large $|V_{ts}^R|$. In this work, we also investigate the CP violating observable, thus, we keep the CP violating phases with the following parametrization,
\be\label{VR}
V_{CKM}^R=\left(\begin{array}{ccc} e^{-i \phi_1} & 0 & 0 \\  0 & c_{23}e^{-i\phi1} & -s_{23} e^{i(\phi_2-\phi_1)}  \\ 0 & s_{23}e^{-i\phi_2} & c_{23} \end{array}\right),
\ed
where $s_{23}\equiv\sin\theta_{23}$, $c_{23}\equiv\cos\theta_{23}$, $\theta_{23}$ represents the mixing angle between the second and the third generations of quarks.
This formula is also a simple version of \cite{Blanke:2011ry} in the limit of $\theta_{12}=\theta_{13}=0$.

So far, we have briefly introduced the LRSM. We summarize the free parameters used in this work: (i) the mass of heavy charged gauge boson $W_2^\pm$, and the masses of heavy charged and neutral Higgs bosons,
\be
M_{W_2},~~~M_H,
\ed
(ii) the right-handed CKM matrix elements, $V_{CKM}^R$, in our formula, they are
\be
\theta_{23},~~~\phi_1,~~~\phi_2,
\ed
(iii) the ratio of right- and left-handed gauge coupling constants,
\be g_R/g_L,\ed
(iv) the ratio of VEVs in the scalar bidoublet $\phi$,
\be \tan\beta,\ed
and (v) the spontaneous CP violating phase $\omega$. In this work, one will find that the observables we used are not very sensitive to $\omega$, therefore we will fix it to be $\omega=0$ in the discussions for simplicity.

\section{The $C_{7\gamma}$ and $C_{7\gamma}^{\prime}$ in the LRSM}
The effective Hamiltonian of $b\to s\ga$ is \cite{Buchalla:1995vs}
{
\begin{equation}
\mathcal{H}_{\rm eff} = -{4G_F\over\sqrt{2}}V_{ts}^{L*}V_{tb}^L \sum_{i=1}^{8} \left[ C_{i}(\mu) Q_{i}(\mu)+C_{i}^{\prime}(\mu) Q_{i}^{\prime}(\mu)\right],
\end{equation}
where the operators $Q_{i}^{\prime}$ are obtained from $Q_{i}$ by replacing $P_{L}\leftrightarrow P_{R}$.
The leading contribution to the $b\to s\gamma$ decay is from the operators of
\be
Q_{7\gamma}={e \over16\pi^2}m_b\bar s\sigma^{\mu\nu}P_Rb F_{\mu\nu},~~~~
Q_{7\gamma}^\prime={e \over16\pi^2}m_b\bar s\sigma^{\mu\nu}P_Lb F_{\mu\nu}.
\ed
We can find that the polarization of the photon radiated from $Q_{7\gamma}$ is left-handed, but right-handed from $Q_{7\gamma}^{\prime}$. In the SM, $C_{7\gamma}^{\prime}/C_{7\gamma}\approx m_{s}/m_{b}\ll 1$, and hence the photon polarization is dominantly left-handed. To test the $wrong$ polarization in the LRSM, we will study  $C_{7\gamma}^{\prime}/C_{7\gamma}$ in this work.

Note that the photon polarization is not directly related to the ratio of $C_{7\gamma}^{\prime}/C_{7\gamma}$, but can also be affected by other operators. It was discussed in \cite{Grinstein:2004uu} that the wrong polarization in the SM could be as large as ten percent contributed by the operator of $Q_{2}= \bar c \gamma_{\mu} P_{L} b \bar s \gamma^{\mu} P_{L} c$.  Therefore, the new physics effects on the polarization could be determined in experiments only if the wrong polarization was larger than ten percent. Besides, the LRSM contributions to $Q_{2}$ and $Q_{2}^{\prime}$ are negligible due to the large mass of heavier charged gauge bosons. Hence, within the leading effects, we investigate the ratio of $C_{7\gamma}^{\prime}/C_{7\gamma}$ to study the wrong polarization in the following discussions.
}

In the LRSM, the penguin diagrams with charged gauge bosons $W_1^\pm$, $W_2^\pm$ and charged Higgs $H^\pm$ in the loops, can contribute to the $b\to s\ga$ processes.
The full formulae of $C_{7\gamma}$ and $C_{7\gamma}^\prime$ are shown in Appendix \ref{app:C7}.
Using the approximation of expansion to the order of $\mathcal{O}(\epsilon^4)$ for $C_{7\ga}$, the Wilson coefficients are
\be\label{c7lw}
&&C_{7\gamma}(\mu_{W_1})_{W_1}={1\over2}\bigg[\left(1-{g_R^2\over g_L^2}{M_{W_1}^4\over M_{W_2}^4}\sin^22\beta\right)A_{\rm SM}(x_t)
\\
&&\ \ \ +{m_t\over m_b}{g_R^2\over g_L^2}{V_{tb}^{R}\over V_{tb}^L}{M_{W_1}^2\over M_{W_2}^2}\sin2\beta e^{i w}\left(\Big(1+{g_R^2\over g_L^2}{M_{W_1}^2\over M_{W_2}^2}\sin^22\beta\Big)A_{\rm LR}(x_t)-{M_{W_1}^2\over M_{W_2}^2}A_{\rm LR}(\tilde x_t)\right)\non\\
&&\ \ \ +{ {m_c\over m_b}{g_R^2\over g_L^2}{V_{cs}^{L*}V_{cb}^{R}\over V_{ts}^{L*}V_{tb}^L}{M_{W_1}^2\over M_{W_2}^2}\sin2\beta e^{i w}\left(\Big(1+{g_R^2\over g_L^2}{M_{W_1}^2\over M_{W_2}^2}\sin^22\beta\Big)A_{\rm LR}(x_c)-{M_{W_1}^2\over M_{W_2}^2}A_{\rm LR}(\tilde x_c)\right)}\bigg],
\non\\
&&C_{7\gamma}(\mu_H)_{H^\pm}={1\over2}\left[{m_t\over m_b}{\tan2\beta\over\cos2\beta}e^{i w}{V_{tb}^R\over V_{tb}^L}A_H^1(y)+\tan^22\beta A_H^2(y)\right],\label{c7lh}
\\
&&C_{7\gamma}^\prime(\mu_{W_1})_{W_1}={1\over2}{m_t\over m_b}{g_R^2\over g_L^2}{V_{ts}^{R*}\over V_{ts}^{L*}}{M_{W_1}^2\over M_{W_2}^2}\sin2\beta e^{-i w}A_{\rm LR}(x_t),\label{c7rw}
\\
&&C_{7\gamma}^\prime(\mu_H)_{H^\pm}={1\over2}\left[{m_t\over m_b}{\tan2\beta\over\cos2\beta}e^{-i w}{V_{ts}^{R*}\over V_{ts}^{L*}}A_H^1(y)+{V_{ts}^{R*}V_{tb}^R\over
V_{ts}^{L*}V_{tb}^L}{1\over\cos^22\beta} A_H^2(y)\right],\label{c7rh}
\ed
where $x_i=m_i^2/M_{W_1}^2$, $\tilde x_i=m_i^2/M_{W_2}^2$, $i=c,t$, $y=m_t^2/M_{H}^2$, and the loop functions are shown in Appendix \ref{app:C7}. The first term in Eq.(\ref{c7lw}) is the SM contribution.
Since the contribution of $C_{7\ga}^\prime$ to branching ratio is of order $\mathcal O(\epsilon^4)$, we have to consider the LRSM contribution up to the order of $\mathcal O(\epsilon^4)$ for $C_{7\ga}$ in which the contributions from LRSM and SM are interfered with each other.
We neglect the contributions from the penguin diagrams with $W_2$ in the loops since they are at a higher order than $\epsilon^4$.
 Note that the dependence on the $W_2$ mass of Eqs. (\ref{c7lw},~\ref{c7rw}) is from the mixing  between $W_L^\pm$ and $W_{R}^\pm$ in $W_1^\pm$ in Eq.(\ref{eq:sinzeta}).
In this work, the contribution from charged Higgs is also negligible unless $\tan\beta\sim1$, because we choose its mass to be higher than several TeV.
The loop functions $A_H^{1,2}(y)$ are two or three orders smaller than $A_{SM}(x_1^t)$ or $A_{LR}(x_1^t)$.
Besides, the charm contribution cannot be neglected in the LRSM as explained in the Appendix \ref{app:C7}.
\begin{figure}[tpbh]
\begin{center}
\includegraphics[scale=0.27]{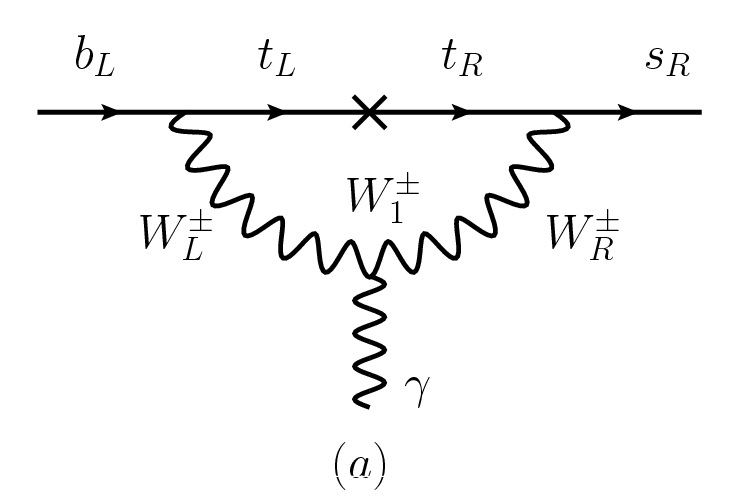}~~~~~
\includegraphics[scale=0.3]{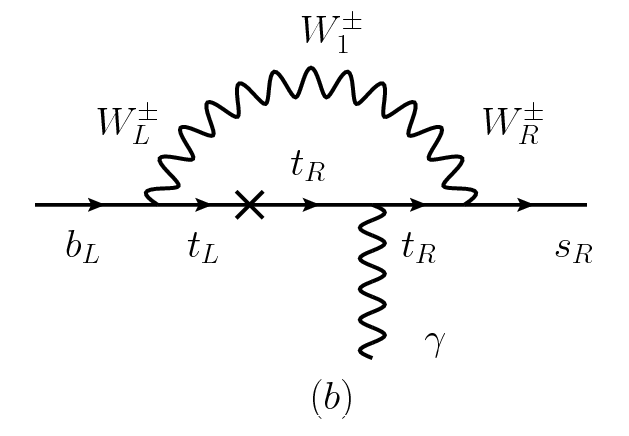}
\caption{The dominant contribution to $b_L\to s_R\ga_R$ in the Left-Right Symmetric Model.  The cross on the quark line indicates a helicity flip via the top quark mass term.
In the LRSM, this process leading {\it wrong helicity} can be enhanced with respect to the SM  by two factors: i) $m_t/m_b$ due to  the helicity flip in the internal top quark line, and ii)  $|V_{ts}^R/V_{ts}^L|$ at the vertex $W_Rt_Rs_R$ if we allow a large value for $|V_{ts}^R|$. }\label{Feyndiag}
\end{center}
\end{figure}

The dominant contribution from the LRSM is $C_{7\gamma}^\prime(\mu_W)_{W_1}$ in Eq.(\ref{c7rw}), from the penguin diagram with $W_1^\pm$ and top quark in the loop, in which the $W_L^\pm$ couples to $b$ quark and $W_{R}^\pm$ couples to $s$ quark, as shown in Fig.\ref{Feyndiag}.
It is enhanced by the factor of $m_t/m_b$ since the helicity flip occurs in the internal top quark line which is resulted by the mass eigenstate $W_1^\pm$ as a mixture of $W_L^\pm$ and $W_R^\pm$.
Compared to the SM contribution, another effect of enhancement is the factor of $|{V_{ts}^{R}/ V_{ts}^{L}}|$ from the vertex of $W_Rt_Rs_R$.
Thus, the value of $V_{ts}^R$ is the key element to have a large $C_{7\gamma}^\prime/C_{7\gamma}$.
If we allow a large value of $|V_{ts}^R|$, this factor of enhancement could be as large as $1/\lambda^2$, where $\lambda\approx0.2$ is one of the parameters in the Wolfenstein formula of left-handed CKM matrix.
In \cite{Bertolini:2012pu}, the authors shows that the mechanism of enhancement of $C_{7\gamma}^\prime/C_{7\gamma}$ in the LRSM is the same as the chromomagnetic one.

\section{Meson mixings in LRSM}\label{sec:Mesonmixing}
The effective Hamiltonian for $\Delta F=2$ transitions can be
written in a general form  \cite{Buras:2000if,Buras:2001ra}
 \be
 \mathcal{H}_{\rm eff}^{\Delta F=2}={G_F^2M_{W_1}^2\over16\pi^2}\sum_{i}
 C_i(\mu)Q_i(\mu),
 \ed
where the CKM matrix elements are involved in the Wilson coefficients $C_i$. Here we factored out $M_{W_1}$ for convenience.
 There are 8 operators responsible for $M^0-\overline M^0$ ($M^0=K,B_d,B_s$)
oscillations \cite{Buras:2000if} \cite{Buras:2001ra}, taking Kaon mixing as example,
 \be\label{de:opMMmix}
 Q_1^{\rm VLL}(K)&=&(\bar s^\alpha\ga_\mu P_Ld^\alpha)(\bar s^\beta\ga^\mu
 P_Ld^\beta),~~~~~
  Q_1^{\rm VRR}(K)=(\bar s^\alpha\ga_\mu P_Rd^\alpha)(\bar s^\beta\ga^\mu
 P_Rd^\beta),
 \non\\
 Q_1^{\rm LR}(K)&=&(\bar s^\alpha\ga_\mu P_Ld^\alpha)(\bar s^\beta\ga^\mu
 P_Rd^\beta),
  \non\\
 Q_2^{\rm LR}(K)&=&(\bar s^\alpha P_Ld^\alpha)(\bar s^\beta
 P_Rd^\beta),\\
 Q_1^{\rm SLL}(K)&=&(\bar s^\alpha P_Ld^\alpha)(\bar s^\beta
 P_Ld^\beta),~~~~~~~~~~~~
  Q_1^{\rm SRR}(K)=(\bar s^\alpha P_Rd^\alpha)(\bar s^\beta
 P_Rd^\beta),
 \non\\
  Q_2^{\rm SLL}(K)&=&(\bar s^\alpha \sigma_{\mu\nu}P_Ld^\alpha)(\bar s^\beta
  \sigma^{\mu\nu} P_Ld^\beta),~~~
  Q_2^{\rm SRR}(K)=(\bar s^\alpha\sigma_{\mu\nu}P_Rd^\alpha)(\bar s^\beta
 \sigma^{\mu\nu}P_Rd^\beta),\non
 \ed
where $\sigma_{\mu\nu}={1\over2}[\ga_\mu,\ga_\nu]$ and
$P_{L,R}={1\over2}(1\mp\ga_5)$. We can readily write the corresponding operators for the $B_d^0$ or $B_s^0$ mixings by replacing $s\to b$ and $d\to d $ or $s$, respectively. In the SM, only the one operator,
$Q_{1}^{\rm VLL}$, contributes to this process.

The amplitude for $M-\overline M$ mixing is then
 \be\label{ampMM}
 \mathcal A(M\to\overline M)={G_F^2M_{W_1}^2\over16\pi^2}\sum_{i,a}
 C_i^a(\mu)\la\overline M|Q_i^a(\mu)|M\ra.
 \ed
The hadronic matrix elements of these operators are shown in  \cite{Buras:2001ra}, in case of $K^0-\overline K^0$ mixing,
 \be\label{de:mem}
\la Q_1^{\rm VLL}(\mu)\ra={2\over3}m_K^2 F_K^2 B_1^{\rm VLL}(\mu),&&
\non\\
\la Q_1^{\rm LR}(\mu)\ra=-{1\over3}R(\mu)m_K^2 F_K^2 B_1^{\rm LR}(\mu),&&
~~~\la Q_2^{\rm LR}(\mu)\ra={1\over2}R(\mu)m_K^2 F_K^2 B_2^{\rm LR}(\mu),
\\\la Q_1^{\rm SLL}(\mu)\ra=-{5\over12}R(\mu)m_K^2 F_K^2 B_1^{\rm SLL}(\mu),&&
~~~\la Q_2^{\rm SLL}(\mu)\ra=-R(\mu)m_K^2 F_K^2 B_2^{\rm SLL}(\mu),\non
\ed
where
\be\label{eq:chiralenhance}
R(\mu)=\left({m_K\over m_s(\mu)+m_d(\mu)}\right)^2
\ed
is the chiral enhancement factor and $B_i$ are bag parameters which is given in Appendix~\ref{app:bag}. The hadronic matrix elements of the operators $VRR$ and $SRR$ are the same as their corresponding $VLL$ and $SLL$ ones because the chiral conservation of QCD.
{  It should be noted that the $K^0-\overline K^0$ system provides the most severe constraints on new physics models, due to the large enhancement factor $1/m_s$ in $R(\mu)_K$  comparing to the $B_{d,s}$ systems.}

At the high energy scale in the LRSM, there are three kinds of contributions to the $\Delta F=2$ processes: box diagrams with two charged gauge bosons $W_{1,2}^{\pm}$, box diagrams with charged Higgs bosons $H^{\pm}$, and tree diagram with flavor-changing neutral Higgs bosons.
We found that the contributions from box diagrams with charged Higgs bosons are always smaller than those with gauge bosons, because of $M_{H}> M_{W_2}$. Therefore they are neglected in this work.

Now, let us see the box diagram with $W_{1,2}^\pm$. The Wilson coefficients for the box diagrams with two $W_1$ (i.e. SM contribution), one $W_1^\pm$ and one $W_2^\pm$, and two $W_2^\pm$ are written respectively as
 \be\label{Kbox}
\Delta_{\rm gauge}C_1^{\rm VLL}(\mu_{W_1})&=& \sum_{i,j=c,t}\lambda_i^{LL}\lambda_j^{LL}S_{LL}(x_i,x_j),
\\
\Delta_{\rm gauge}C_2^{\rm LR}(\mu_{W_2})&=& \sum_{i,j=u,c,t}\lambda_i^{LR}\lambda_j^{RL}S_{LR}(x_i,x_j,\eta),
\\
\Delta_{\rm gauge}C_1^{\rm VRR}(\mu_{W_2})&=& \sum_{i,j=c,t}\lambda_i^{RR}\lambda_j^{RR}S_{RR}(\tilde x_i,\tilde x_j),
\ed
where
$\lambda^{AB}_i=V_{iq}^{A*}V_{iq'}^{B}$, $q,q'=d,s,b$, $A,B=L,R$, $x_i=m_i^2/M_{W_1}^2$, $\tilde x_i=m_i^2/M_{W_2}^2$, $i=u,c,t$, $\eta=M_{W_1}^2/M_{W_2}^2$. In the $LL$ and $RR$ sectors, the GIM mechanism has been used and then the contribution from $\lambda_u$ is eliminated, while in the $LR$ sector, the GIM mechanism is invalid because of the combination of left- and right-handed CKM matrix elements.
The loop functions can be found in the Appendix \ref{app:box}. We find that $\Delta_{\rm gauge}C_2^{\rm LR}$ dominates the LRSM contribution, while $\Delta_{\rm gauge}C_1^{\rm VRR}$ is negligible due to the high mass of $W_2^\pm$.
Our results agree with those in \cite{Zhang:2007da,Blanke:2011ry,Maiezza:2010ic}. Considering the right-handed CKM matrix in (\ref{VR}) with some elements as zero, the dominant contributions to the box diagrams to Kaon, $B_d$ and $B_s$ mixing yeild:
\be\label{eq:kbox}
\Delta C_2^{LR}(\mu_{W_2})^K_{Box}
&=& V_{us}^{L*}V_{ud}^RV_{cs}^{R*}V_{cd}^L S_{LR}(x_u,x_c,\eta)
+V_{us}^{L*}V_{ud}^RV_{ts}^{R*}V_{td}^L S_{LR}(x_u,x_t,\eta),
\ed
\be\label{eq:bdbox}
\Delta C_2^{LR}(\mu_{W_2})^{B_d}_{Box}
&=& V_{ub}^{L*}V_{ud}^RV_{cb}^{R*}V_{cd}^L S_{LR}(x_u,x_c,\eta)
+V_{ub}^{L*}V_{ud}^RV_{tb}^{R*}V_{td}^L S_{LR}(x_u,x_t,\eta),
\ed
\be\label{eq:bsbox}
\Delta C_2^{LR}(\mu_{W_2})^{B_s}_{Box}
&=& V_{tb}^{L*}V_{ts}^RV_{cb}^{R*}V_{cs}^L S_{LR}(x_c,x_t,\eta)
+V_{tb}^{L*}V_{ts}^RV_{tb}^{R*}V_{ts}^L S_{LR}(x_t,x_t,\eta),
\ed
Due to $V_{cd}^R=V_{td}^R=0$ in (\ref{VR}), only the box diagrams with at
least one up quark contribute to the Kaon and $B_d$ mixing, as shown in Eqs.(\ref{eq:kbox}) and (\ref{eq:bdbox}).
Although the loop functions with an up quark in the box diagrams are numerically small, we consider this contribution to the Kaon mixing system because of the effect of enhancement in Eq.(\ref{eq:chiralenhance}).
On the other hand, such enhancement is absent for the $B_d^0-\overline B_d^0$ mixing.
Thus, together with the suppression by $V_{ub}^R$, we neglect the contribution from box diagrams to $B_d^0-\overline B_d^0$ oscillation.
For the system of $B_s^0-\overline B_s^0$ mixing, the dominant contributions in box diagrams are from the charm quark or top quark in the loops.
In Eq.(\ref{eq:bsbox}), we neglect the small contribution from two charm quarks in the box diagram because of the small value of the loop function.
One more diagram with one charm quark and one top quark is also neglected due to the suppression of corresponding left-handed CKM matrix elements.

The tree-level contributions from the flavor-changing neutral Higgs (NH) are expected to dominate the $\Delta F=2$ processes. The corresponding Wilson coefficient is
 \be
 \Delta_{\rm NH}C_2^{LR}(\mu_H)=-{128\pi^2\over \sqrt{2}M_H^2G_F}\sum_{i,j=u,c,t}\lambda_i^{LR}\lambda_j^{RL}\sqrt{x_i(\mu_H)x_j(\mu_H)},
\ed
where the up-type quark masses are evaluated at the scale of $\mu_H$ due to the renormalization of Yukawa couplings.
Because of $V_{us}^R=V_{cd}^R=V_{td}^R=0$ in our right-handed CKM matrix elements in (\ref{VR}), similarly to the case of box diagrams, only two terms contribute to Kaon mixing, and $B_d^0$ mixing.
Due to the tiny mass of up quark, the flavor-changing neutral Higgs contribution in the LRSM to $B_d^0-\overline B_d^0$ system is much smaller than the SM one, thus it can be neglected.
Thereby we will not consider the $B_d^0-\overline B_d^0$ mixing in the following discussion in this work.
For the Kaon mixing, the LRSM effects do not change the mass difference very much. However, since the indirect CP violation $\epsilon_K$ is very small in the SM, the LRSM contribution to  $\epsilon_K$ has to be examined very carefully.
The flavor-changing neutral Higgs (NH) contributions to the $K^0-\overline K^0$ and $B_s^0-\overline B_s^0$ mixing are
\be
\Delta C_2^{LR}(\mu_{H})^K_{NH}&=&{-128\pi^2\over \sqrt{2}M_{W_1}^2G_F M_H^2}\Big(m_u(\mu_H)m_c(\mu_H)V_{us}^{L*}V_{ud}^RV_{cs}^{R*}V_{cd}^L
\non\\
&&+m_u(\mu_H)m_t(\mu_H)V_{us}^{L*}V_{ud}^RV_{ts}^{R*}V_{td}^L\Big),
\ed
\be
\Delta C_2^{LR}(\mu_{H})^{B_s}_{NH}&=&{-128\pi^2\over \sqrt{2}M_{W_1}^2G_F M_H^2}\Big(m_c(\mu_H)m_t(\mu_H)V_{tb}^{L*}V_{ts}^RV_{cb}^{R*}V_{cs}^L
\non\\&&+m_t(\mu_H)^2V_{tb}^{L*}V_{ts}^RV_{tb}^{R*}V_{ts}^L\Big),
\ed
where the up-type quark masses are evaluated at the scale of $\mu_H$.
With the same argument to the case of box diagrams above, we neglect two terms in the $B_s^0-\overline B_s^0$ mixing.
As the quark masses don't change too much at $\mu_H=(20\sim50)$TeV, we fix them as
\be
m_u(\mu_H)=0.001GeV,~~~m_c(\mu_H)=0.5GeV,~~~m_t(\mu_H)=120GeV.
\ed

\section{Numerical Analysis}
\subsection{$C_{7\gamma}^{\prime}/C_{7\gamma}$}
$C_{7\gamma}$ and $C_{7\gamma}^{\prime}$ are mainly constrained by the branching ratio of $B\to X_s\ga$ whose current world average value is\cite{HFAG}
 \be
Br(B\to X_s\gamma)^{\rm
exp}_{E_{\gamma}>1.6GeV}=(3.43\pm0.22)\times10^{-4}.
 \ed
 The SM prediction within next-to-next-to-leading order of QCD is
\cite{Misiak:2006zs}
 \be
Br(B\to X_s\gamma)^{\rm
SM}_{E_{\gamma}>1.6GeV}=(3.15\pm0.23)\times10^{-4},
 \ed
Adding the LRSM contributions, the $C_{7\gamma}^{\rm SM}$ term is extended to:
\be
C_{7\gamma}(\mu_b)=C_{7\gamma}^{\rm SM}(\mu_b)+C_{7\gamma}^{\rm NP}(\mu_b),~~~~~C_{7\gamma}^\prime(\mu_b)=C_{7\gamma}^{\prime {\rm NP}}(\mu_b),
\ed
where the  components of NP terms are already expressed in the previous section.
Note that the left-handed and the right-handed polarization terms, i.e., $C_{7\gamma}$ and $C_{7\gamma}^{\prime}$, do not interfere with each other.
In the branching ratio of $B\to X_s\gamma$ in the SM, the contribution from tree level operator, i.e. $C_2^{\rm SM}$ is very large. Therefore, the overlap of $C_{7\gamma}^{(\prime){\rm NP}}$ and the $C_2^{\rm SM}$ is not negligible.
On the other hand,
the new tree-level contribution in the LRSM (see  Eq.~(\ref{appA:tree}) in Appendix~\ref{app:Lagrangian} for the full expression) turns out to be less than a few percent of $C_2^{\rm SM}$ due to the high mass of the $W_2^{\pm}$.
For the $LR$ and $RL$ operators, they could contribute with the chirality flip by the charm quark mass which is much smaller than top quark mass.
Thus, the tree level contributions from the LRSM can be safely neglected.

While we use the full equation in our numerical study, we show a simplified formula using typical input parameters for illustration. The branching ratio of $b\to s\ga$ can be written in terms of new physics (NP) Wilson coefficients at the scale of $\mu_b=4.6$GeV
\be
&&Br(B\to X_s\ga)_{E_\ga>1.6{\rm GeV}}
\non\\
&&\simeq \bigg(3.15+18.83\left(\left|C_{7\gamma}^{NP}(\mu_b)\right|^2+\left|C_{7\gamma}^{\prime NP}(\mu_b)\right|^2\right)-11.79Re\left[C_{7\gamma}^{NP}(\mu_b)\right]
\bigg)\times10^{-4}.\label{eq:brbsga}
\ed
In this formula, we update the NLO analysis from \cite{Kagan:1998ym} by the new results of quark masses and CKM matrix elements in \cite{PDG}, and by the NNLO SM result for this branching ratio from \cite{Misiak:2006ab,Misiak:2006zs}. For the Wilson coefficients,
we use the leading-order (LO) formulae for the $W_1^\pm$ and $H^\pm$ contributions at the matching scale of $\mu_W$ and $\mu_H$, and then evolve them down to the $\mu_b$ scale separately at the leading logarithm (LL), as is done  in~\cite{Blanke:2011ry}.
The procedures of evolution are the same for $C_{7\gamma}$ and $C_{7\gamma}^\prime$ because QCD is conserved to the left-right symmetry. The mixing between $O_7$ and other operators are considered in the evolution. Besides, $C_{7\gamma}$ and $C_{7\gamma}^\prime$ are actually the effective $C_{7\gamma}^{\rm eff}$ and $C_{7\gamma}^{\prime \rm eff}$, but we use $C_{7\gamma}^{(\prime)}$ in this article for simplicity without ambiguity.

Numerically, $C_{7\ga}^\prime(\mu_b)$ and $C_{7\ga}(\mu_b)$ can be expressed as a function of the parameters in the LRSM:
\be\label{eq:c7lmub}
C_{7\ga}^\prime(\mu_b)&\simeq& 466 {g_R^2\over g_L^2} {M_{W_1}^2\over M_{W_2}^2} \sin2\beta V_{ts}^{R*} e^{-i \omega},
\\
C_{7\ga}(\mu_b)&\simeq& - 0.32+ {g_R^2\over g_L^2}{M_{W_1}^2\over M_{W_2}^2} \sin2\beta\left(7.5 V_{cb}^{R}-18.9 V_{tb}^R\right)e^{i \omega}.
\ed
The first term in $C_{7\ga}$ is from the SM contribution, while the second term in  $C_{7\ga}$ is from  the LRSM contribution which is suppressed by ${M_{W_1}^2/ M_{W_2}^2}$, and can be neglected. So $C_{7\ga}$ can be taken as the SM one in an approximation, $C_{7\ga}\simeq C_{7\ga}^{\rm SM}$.  Then
\be\label{c7poc7num}
{C_{7\ga}^\prime(\mu_b)\over C_{7\ga}(\mu_b)}\simeq -1455 {g_R^2\over g_L^2} {M_{W_1}^2\over M_{W_2}^2} \sin2\beta \sin\theta_{23} e^{i(\phi_2- \omega)}.
\ed
We can find that $|C_{7\gamma}^\prime/C_{7\gamma}|$ could be as large as $0.5$ if $\sin\theta_{23}\sim 1$, $M_{W_2}\sim 2$ TeV, $g_R/g_L\sim 1$ and $\tan\beta\sim 10$.

\subsection{Meson mixings}\label{sec:nmm}
To constrain the LRSM, we use the mass difference $\Delta M_s$ and the CP violating phase $\phi_s$ in $B_s^0-\overline B_s^0$ system, and the indirect CP violation in neutral Kaon mixing system.
The current average values of mass difference and CP violating phase of $B_s^0-\overline B_s^0$ mixing are \cite{HFAG}
 \be
\Delta M_{B_s}&=&17.719\pm0.043~{\rm ps}^{-1},\\
\phi_s^{c\bar cs}&=&-0.013^{+0.083}_{-0.090}.
\ed
The indirect CP violation in neutral Kaon mixing is \cite{PDG}
\be
|\epsilon_K|^{\rm exp}=(2.228\pm0.011)\times10^{-3}.
\ed

The mass difference and the CP violating phase of  $B_{s}^0$ systems are
 \be
\Delta M_{B_s}&=&2|M_{12}^s|,~~~~~\phi_s=-\arg \left(M_{12}^s\right),
\ed
where the matrix element $M_{12}^s$ is defined by
\be
2 m_{B_s} M_{12}^s=\la B_s^0|\mathcal{H}_{\rm eff}|\overline B_s^0\ra.
\ed
{
The LRSM contribution to the tree-level two-body hadronic decays is suppressed by the high mass of $W_2$ and hence negligible. Therefore we take the weak phase difference $\phi_s^{c\bar cs}$ between the $B_s^0-\overline B_s^0$ mixing and the $b\to c\bar cs$ decay amplitudes as the phase in the mixing.
}

The CP violation in the neutral Kaon meson decays is defined as \cite{Buras:2008nn}\cite{Buras:2010pza}
 \be
 \varepsilon_K={\kappa_\epsilon e^{i\varphi_\epsilon}\over
 \sqrt{2}(\Delta M_K)_{\rm exp}}{\rm Im}(M_{12}^K),
 \ed
where $\varphi_\epsilon=(43.51\pm0.05)^\circ$\cite{PDG} and
 $\kappa_\epsilon=0.94\pm0.02$ are corrections due to long-distance
 effect.

Finally, at the $\mu_K$ scale, the matrix element of $M_{12}^K$ is summarized as
\be
M_{12}^K=M_{12}^{SM}(K)+M_{12}^{Box}(K)+M_{12}^{NH}(K),
\ed
where $M_{12}^{SM}(K)$, $M_{12}^{Box}(K)$, $M_{12}^{NH}(K)$ are from the contributions of the SM, box diagrams with $W_2^\pm$ and flavor-changing neutral Higgs bosons, respectively. They can be expressed by the parameters in the LRSM,
\be
M_{12}^{SM}(K)&=&1.58\times10^{-15}-( 1.19\times10^{-17})i,\\
M_{12}^{Box}(K)&=&r^2 \eta\big[-22\ln\eta\cos\theta_{23}\non\\
&&~~~~+(208+97\ln \eta)\sin\theta_{23}e^{i(\phi_2-\phi_1-\beta_{\rm CKM})} \big]\times10^{-17},\\\label{eq:M12Kbox}
M_{12}^{NH}(K)&=&\eta_H\big[(4137-339\ln \eta_{H})\cos\theta_{23}
\non\\
&&~~~~-(41573-3402\ln \eta_{H})\sin\theta_{23}e^{i(\phi_2-\phi_1-\beta_{\rm CKM})} \big]\times10^{-17},\label{eq:M12KNH}
\ed
where $r\equiv g_R/g_L$, $\eta\equiv M_{W_1}^2/M_{W_2}^2$, $\eta_H\equiv M_{W_1}^2/M_H^2$, and $\beta_{\rm CKM}$ is the CP violating phase $\beta$ in the left-handed CKM matrix, to be distinguished by the $\beta$ parameter in the Higgs sector in the LRSM.
We can find that the real part of contribution from the LRSM are much smaller than the SM one, and hence the LRSM would not affect $\Delta M_K$. However, the imaginary part of contributions from the LRSM are in the same order as the SM one, so that $\epsilon_K$ can be significantly affected by the LRSM.

For the $B_s^0-\overline B_s^0$ system, the hadronic matrix element $M_{12}^{B_s}$ can be expressed similarly as $M_{12}^K$,
\be
M_{12}^{SM}(B_s)&=&(55.4+2.0i)\times10^{-13},\\
M_{12}^{Box}(B_s)&=&r^2 \eta\big[(-453-141\ln \eta)\cos\theta_{23}\sin\theta_{23}e^{-i\phi_2}\non\\
&&~~~~~~+(-61-28\ln \eta)\sin^2\theta_{23}e^{i(\phi_1-2\phi_2)} \big]\times10^{-11},\\\label{eq:M12Bsbox}
M_{12}^{NH}(B_s)&=&\eta_H\big[(1466-120\ln \eta_{H})\cos\theta_{23}\sin\theta_{23}e^{-i\phi_2}\non\\
&&~~~~~~+(144-12\ln \eta_{H})\sin^2\theta_{23}e^{i(\phi_1-2\phi_2)} \big]\times10^{-9},\label{eq:M12BsNH}
\ed
where $r\equiv g_R/g_L$, $\eta\equiv M_{W_1}^2/M_{W_2}^2$, and $\eta_H\equiv M_{W_1}^2/M_H^2$.
We can find that the contributions from $W_2^\pm$ and neutral Higgs bosons are at the same order of, or even larger than the SM contribution, within the region of values of parameter we discussed. Then the constraints from $B_s^0-\overline B_s^0$ mixing are severe. The CP violating phases $\phi_1$ and $\phi_2$ become important to provide solutions without fine-tuning.

In summary, concerning $\theta_{23}$, if $\theta_{23}\simeq 0$ (or $\pi$), i.e. $\sin\theta_{23}\simeq0$, we found that all the constraints from the meson mixings can be neglected as long as $M_{W_2}, M_H$ are not quite small, because $M_{12}^{LRSM}(B_s)\propto\sin\theta_{23}$, and the terms with $\cos\theta_{23}$ in $M_{12}^{LRSM}(K)$ are without CP violating phase and hence lose the constraint from $\epsilon_K$. However in order to have a large $|C_{7\gamma}^{\prime}/C_{7\gamma}|$, we need $\sin\theta_{23}$ to be large.
If so, the contributions from the second terms in Eq. (\ref{eq:M12Kbox},\ref{eq:M12Bsbox}) become dominant. Interestingly, these terms depend on the $\phi_1$ and $\phi_2$. Thus, we can find some solutions in terms of $\phi_{1,2}$, which satisfy $M_{12}^{\rm SM}+M_{12}^{\rm Box}+M_{12}^{\rm NH}$ to be within the experimental bounds. Indeed such solutions exist as long as
 $M_H>20$ TeV and/or  $M_{W_2}>2$ TeV.

\subsection{Impacts of the future photon polarization measurement on the LRSM}
In this section, we attempt to estimate how large the $|C_{7\gamma}^{\prime}/C_{7\gamma}|$ could be in the LRSM fully taking into account the various flavour constraints discussed in the previous sections.
As we have already seen, $|C_{7\gamma}^{\prime}/C_{7\gamma}|$
has a simple dependence on the parameters:  it becomes larger for a lower $W_2^\pm$ mass and/or for a larger $\sin\theta_{23}$. However, such a choice of parameter could make  the LRSM contributions to the meson oscillation observables inconsistently large. Thus, we need a careful test of the whole parameter space of ($\theta_{23}, M_{W_2}, M_H, \phi_1, \phi_2$). Let us set our target to find the maximum possible $|C_{7\gamma}^{\prime}/C_{7\gamma}|$ value for given values of $M_{W_2}$ (here, we chose $M_{W_2}=1.5, 2.0, 2.5, 3.0, 3.5, 4.0, 4.5, 5.0$\ TeV), and $M_H$ (we fix it as 20 TeV or 50 TeV).  Then, our first goal is to find a combination of  $\phi_1$ and $\phi_2$ which satisfies all the experimental constraints while keeping $\sin\theta_{23}$ to be large.

It should be mentioned that in order to find that the allowed parameter space from the meson mixing and $B\to X_s\gamma$ branching ratio, we consider not only the experimental errors but also the theoretical errors. The input parameters and their uncertainties taken into account here is listed in Table~\ref{Tab:1}\footnote{The hadronic parameters $B_{2}^{LR}(K)$, $B_1^{VLL}(B)$, $B_2^{LR}(B)$ are calculated by Lattice QCD, but their values are sometimes inconsistent  by different lattice QCD groups, such as in \cite{Bertone:2012cu} and \cite{Boyle:2012qb}, so that there are large uncertainties of these parameters. We take the center values of these parameters from  \cite{Becirevic:2001xt}  for Kaon system and \cite{Bertone:2012cu} for $B_s^0$ system, and take their uncertainty as almost $10\%$. The other $B$ parameters are either determined very well such as $B_{1}^{VLL}(K)$, or do not affect the observables in our work, so we do not consider their uncertainties.
The bag parameters used in this work are shown in Appendix \ref{app:bag}. } .
Some comments on the left-handed CKM matrix are in order. We assume that the $W_R$ contributions to the tree-level processes are negligible, which is reasonable for the large $M_{W_2}$ considered here.
Since $V_{ts}^L$ is extracted by PDG \cite{PDG} from $B_s^0-\overline B_s^0$ oscillation which is affected by the LRSM, we extract $V_{ts}^L$ by using the unitarity relation of the left-handed CKM matrix, so that
$V_{ts}^L=-1/V_{tb}^{L*}(V_{us}^LV_{ub}^{L*}+V_{cs}^LV_{cb}^{L*})$.
The central value and uncertainty of $|V_{tb}^L|$ is taken as the experimental result in \cite{PDG}, which is also from tree-level top quark decays.
$|V_{td}^L|$ and the phase $\beta_{\rm CKM}$ are both determined by the $B_d^0-\bar B_d^0$ oscillations, $|V_{td}^L|$ from $\Delta M_d$, and $\beta_{\rm CKM}$ from CP asymmetry. This process is also box diagrams mediated, but as discussed in the Section~\ref{sec:Mesonmixing}, due to the right-handed CKM matrix elements $V_{cd}^R=V_{td}^R=0$, the effect of the LRSM in $B_d^0$ system can be neglected. So that we could use the values of $\beta_{\rm CKM}$ in \cite{PDG}. Besides, $|V_{td}^L|$ in \cite{PDG} is obtained assuming $|V_{tb}^L|=1$, but we use the experimental result of $|V_{tb}^L|$ in this work, so we normalize $|V_{td}^L|$ by $|V_{tb}^L|$.
\begin{table*}
\begin{center}
\begin{tabular}{|c|c|c|c|}
\hline
\multicolumn{4}{|c|}{hadronic parameters} \\
\hline
 $B_2^{LR}(K)$& $B_{1}^{VLL}(B)$ &  $B_2^{LR}(B)$ & $\eta_{ct}$~\cite{Brod:2010CT}.  \\
\hline
$0.82\pm0.08$ & $0.87\pm0.08 $ & $1.15\pm0.12$&$0.496\pm0.047$ \\
\hline
\hline
\multicolumn{4}{|c|}{CKM parameters~\cite{PDG}} \\
\hline
$|V_{td}^L|$&$|V_{ts}^L|$&$|V_{tb}^L|$&$\beta_{\rm CKM}$ \\
\hline
$(9.44\pm0.83)\times10^{-3}$&$(-46.6\pm3.6)\times10^{-3}$&$0.89\pm0.07$
&$0.373\pm0.013$ \\
\hline
\end{tabular}
\end{center}
\caption{Input parameters  }
\label{Tab:1}
\end{table*}%

Now, we look for combinations of  $\phi_1$ and $\phi_2$ which satisfy all the experimental constraints taking fully into account the uncertainties in these input parameters.
The procedure to find these combinations is described in the appendix~\ref{app:fit}.
We fix the values of $M_{W_2}$ and $M_{H}$ and scan the $(\phi_1, \phi_2)$ space, which is chosen in $[-\pi, \pi]$ with $\pi/50$ interval (total of $10^4$ points for each mass combination). Then, we find the minimum of the $\chi^2$ defined in the appendix~\ref{app:fit}. The number of points which did not exceed any of the experimental constraints or the input parameter ranges are (for $M_{W_2}=$\ (1.5, 2.0, 2.5, 3.0, 3.5, 4.0, 4.5, 5.0, 5.5) TeV, respectively):
\begin{eqnarray}
& (0.01, 0.8,  0.8,  0.9,  0.9, 1.0,  1.1, 1.1, 1.2)\times 10^3 \ {\rm points\ for\ } M_H=20\ {\rm TeV} &, \\
& (0.06, 1.4, 1.8, 2.6, 3.1, 4.2, 5.2, 5.7, 6.2)\times 10^3 \ {\rm points\ for\ } M_H=50\ {\rm TeV} &.
\end{eqnarray}
As we have mentioned in the previous sections, $M_{W_2}\simeq 1.5$ TeV is allowed only for a fine-tuned parameter combinations while for $M_{W_2}> 2.0$ TeV, we find sizable number of allowed points.

\begin{figure}[thpb]\begin{center}
\includegraphics[scale=0.4]{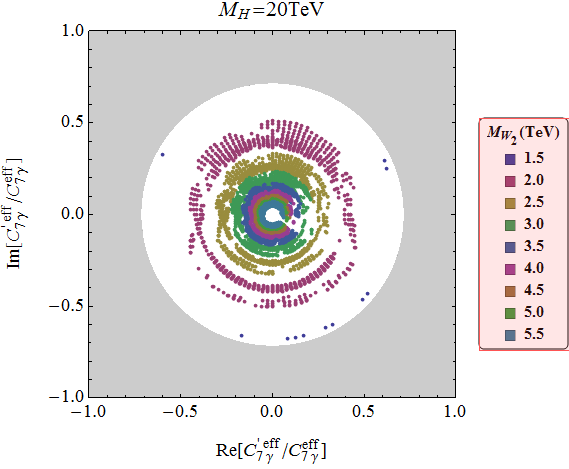}~~~~~~~~~
\includegraphics[scale=0.4]{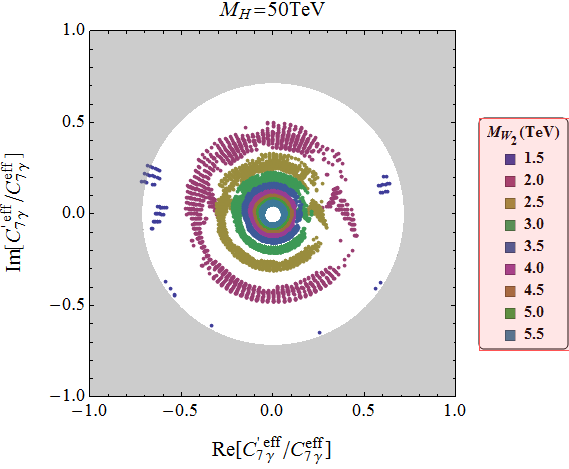}
\caption{Plots of real part and imaginary part of $C_{7\gamma}^{\prime}/C_{7\gamma}$ in the LRSM. {  The left and right figures correspond to the cases that the heavy Higgs mass is 20 and 50 TeV respectively. The white circle represents the constraint from the measured branching ratio of $B\to X_s\gamma$ with three standard deviation, in the scenario assuming $C_{7\gamma}^{NP}=0$, i.e. $C_{7\gamma}=C_{7\gamma}^{SM}$, $C_{7\gamma}^{\prime NP}\in\mathbb{C}$.  } The points from the outside to the inside with different colors represent the cases that the mass of $W_2^\pm$ is taken to be 1.5, 2, 2.5, 3, 3.5, 4, 4.5, 5 and 5.5 TeV respectively, with the circles from the outside to the inside in the figure.  The points represent the solutions that are mainly constrained by $\epsilon_K$, $\Delta M_s$, $\phi_s$ and $Br(B\to X_s\gamma)$.  $|C_{7\gamma}'/C_{7\gamma}|$ would be larger as the mass of $W_2$ decreases. We can find that $|C_{7\gamma}'/C_{7\gamma}|$ can be as large as 0.5 for $M_{W_2}=2$ TeV, and 0.35 for $M_{W_2}=2.5$ TeV.
}\label{fig:ReImC7}
\end{center}\end{figure}

Using these allowed parameter points, we arrive to compute $|C_{7\gamma}^{\prime}/C_{7\gamma}|$. All other parameters are optimized to the values which give the minimum of $\chi^2$ described in the appendix~\ref{app:fit}.  The LRSM corresponds approximately to the scenario II of~\cite{Becirevic:2012dx} where $C_{7\ga}^{NP}\simeq 0$, $C_{7\ga}^{\prime(NP)}\in\mathbb{C}$ ({\it c.f.}  Eq.~(\ref{eq:c7lmub})).  Then we could investigate the effect of the LRSM in a plan of $Re[C_{7\ga}^\prime/C_{7\ga}]$ and $Im[C_{7\ga}^\prime/C_{7\ga}]$, as shown in Figure \ref{fig:ReImC7}.
The left and right figures correspond to the cases that the heavy Higgs mass is 20 and 50 TeV respectively.
The white circle represents the constraint from the measured branching ratio of $B\to X_s\gamma$ with three standard deviation.
The points with different colors represent the cases that the mass of $W_2$ is taken to be 1.5, 2.0, 2.5, 3.0, 3.5, 4.0, 4.5, 5.0 and 5.5 TeV respectively, with the circles from the outside to the inside in the figure.
The points represents the solutions that are mainly constrained by $\epsilon_K$, $\Delta M_s$, $\phi_s$ and $Br(B\to X_s\gamma)$.  $|C_{7\gamma}'/C_{7\gamma}|$ would be larger as the mass of $W_2$ decreases.
One can see that $|C_{7\gamma}^{\prime}/C_{7\gamma}|$ can reach to $\sim (0.5, 0.35, 0.25, 0.15, 0.1)$ for $W_2$ being (2.0, 2.5, 3.0, 3.5, 4.0) TeV.

Finally, we discuss briefly  the discovery potential of the LRSM by the future photon polarization measurements.
There have been various proposals for the experimental method to determine the photon polarization of the $b\to s\gamma$. Here, following the discussion in~\cite{Becirevic:2012dx}, we study the three types of methods; i) The time-dependent $CP$-asymmetry of $B\to K_S\pi^0\gamma$ ($S_{K_S\pi^0\gamma}$),
ii) The angular distribution of the $\overline B\to \overline K_1\ga\to K\pi\pi\ga$ decay ($\lambda_\ga$), and iii) The angular analysis of $\overline B\to \overline K^*(\to K\pi)\ell^+\ell^-$  ($A_T^{(2)}, A_T^{(im)}$).
Detailed descriptions of these observables are given in Appendix~\ref{appG}.
\begin{description}
\item{i)} In the LRSM, $S_{K_S\pi^0\gamma}$ can be written as
\be
S_{K_S\pi^0\ga}\simeq{-2Re\left[{C_{7\ga}^\prime/ C_{7\ga}}\right]\sin2\beta_{\rm CKM}+ 2Im\left[{C_{7\ga}^\prime/ C_{7\ga}}\right]\cos2\beta_{\rm CKM} \over 1+ Re\left[{C_{7\ga}^\prime/ C_{7\ga}}\right]^2+Im\left[{C_{7\ga}^\prime/ C_{7\ga}}\right]^2},
\ed
where $\beta_{\rm CKM}$ is the CP violating phase from the left-handed CKM matrix.
This formula is obtained by assuming $C_{7\ga}^{NP}\simeq0$ and also absence of new physics contributions to the $B_d^0-\overline B_{d}^0$ mixing, which have been derived in the previous sections.
In Figure.3(a), we overlap the curve representing the center value, 1$\sigma$,
2$\sigma$, 3$\sigma$, of the experimental result of $S_{K_S\pi^0\gamma}$ on top of the possible LRSM effects obtained in  Figure 2. We can find that $S_{K_S\pi^0\ga}$ is approximately a linear combination of $Re\left[{C_{7\ga}^\prime/ C_{7\ga}}\right]$ and $Im\left[{C_{7\ga}^\prime/ C_{7\ga}}\right]$.
\item{ii)} In the LRSM, the polarization parameter $\lambda_\gamma$ can be written as
\be
\lambda_\ga={Re\left[{C_{7\ga}^\prime/ C_{7\ga}}\right]^2+Im\left[{C_{7\ga}^\prime/ C_{7\ga}}\right]^2-1\over Re\left[{C_{7\ga}^\prime/ C_{7\ga}}\right]^2+Im\left[{C_{7\ga}^\prime/ C_{7\ga}}\right]^2+1}.
\ed
The future sensitivity of $\lambda_\gamma$ is shown in Figure.3(b). The $\lambda_\ga$ constraint is  circle and insensitive to the complex phases. The SM prediction stands at the central point. As stated in \cite{Becirevic:2012dx}, in the region inside the circle of  $\lambda_\ga\simeq-0.8$, i.e. $\left|C_{7\ga}^\prime /C_{7\ga}\right|\simeq0.3$  we cannot distinguish the LRSM contribution from the SM one. { Therefore, if the discovery of photon polarization using this method in experiments is distinguished from the SM, we can conclude that the mass of $W_2^\pm$ is smaller than 2.5 TeV in case of $g_R/g_L\sim 1$ and $\tan\beta\sim 10$.}
\item{iii)} In the LRSM, the transverse asymmetries can be written as
\be
A_T^{(2)}(0)&=&{2Re\left[{C_{7\ga}^\prime/ C_{7\ga}}\right]\over 1+Re\left[{C_{7\ga}^\prime/ C_{7\ga}}\right]^2+Im\left[{C_{7\ga}^\prime/ C_{7\ga}}\right]^2},\\
A_T^{(im)}(0)&=&{-2Im\left[{C_{7\ga}^\prime/ C_{7\ga}}\right]\over 1+Re\left[{C_{7\ga}^\prime/ C_{7\ga}}\right]^2+Im\left[{C_{7\ga}^\prime/ C_{7\ga}}\right]^2}.
\ed
These two observables can combine together to determine the magnitude and phase of ${C_{7\ga}^\prime/ C_{7\ga}}$, shown in Figure.3(c) and 3(d).

\end{description}

\begin{figure}[thpb]\label{fig:futuremeasure}
\begin{center}
\begin{tabular}{cc}
\includegraphics[scale=0.4]{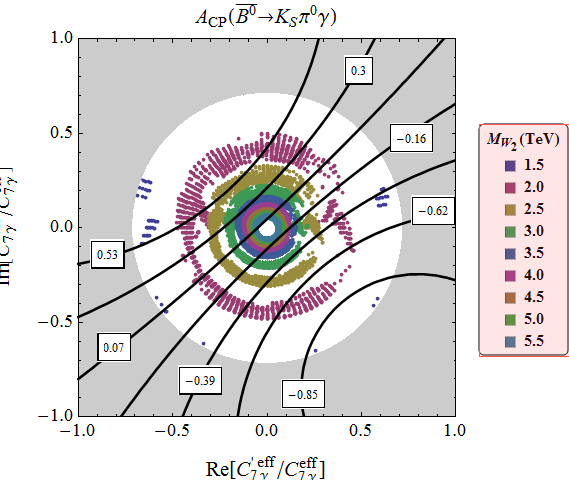}&
\includegraphics[scale=0.4]{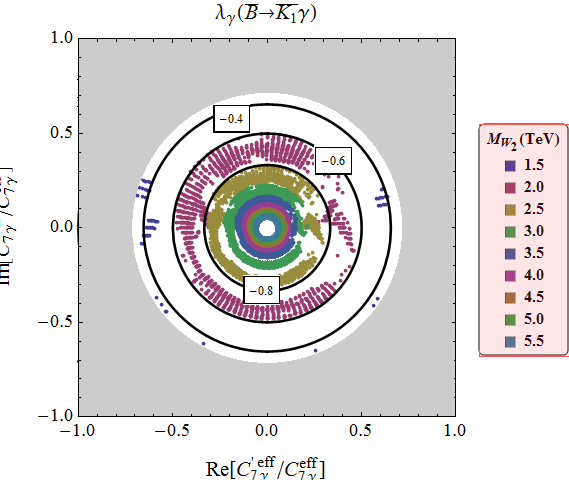}\\
(a)&(b)\\
\includegraphics[scale=0.4]{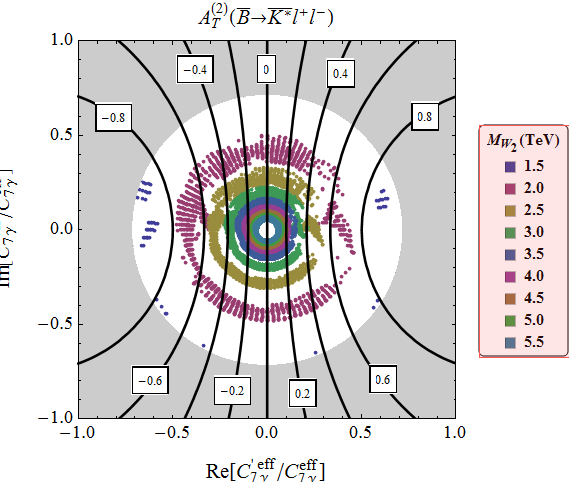}&
\includegraphics[scale=0.4]{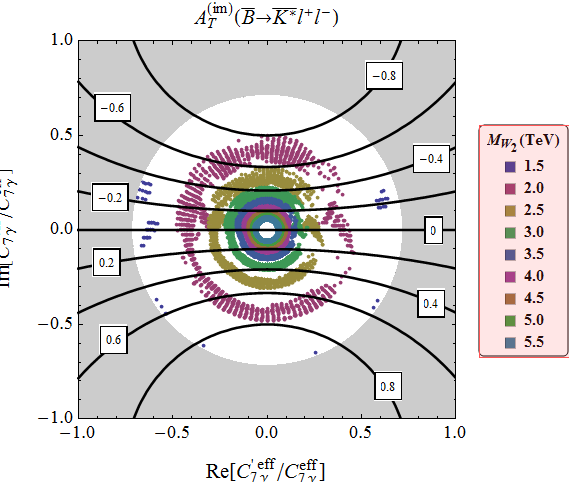}\\
(c)&(d)\\
\end{tabular}
\caption{Prospect of the future constraints on $C_{7\ga}^\prime/C_{7\ga}$ in the LRSM. The four figures (a,b,c,d) correspond to $S_{K_S\pi^0\ga}$, $\lambda_\ga$, $A_T^{(0)}$  and $A_T^{(im)}$ respectively, allowed by a $\pm3\sigma$ error of $Br(B\to X_s\ga)_{\rm exp}$. The points in the figures are the same as those in Fig.2(b) with $M_H=50$TeV. The figure (a) is our result for $S_{K_S\pi^0\ga}$ in a range including $\pm 3\sigma$.  The uncertainty of measurement for $S_{K_S\pi^0\ga}$ can be reduced down to 2$\%$ in the future Belle II experiment, in which the contours will be 10 times narrow than those in figure (a). In figures (b,c,d), the contours are the uncertainty for $\lambda_\ga$, $A_T^{(0)}$  and $A_T^{(im)}$ in the future measurements at LHCb and Belle II, which can be achieved with the errors as small as 0.2.}
\end{center}
\end{figure}

\subsection{Complementary to the direct search at LHC}

The direct search of $W_R$ or $W^\prime$, which also provides an important constraint on the LRSM,  is carried out at the high energy colliders, such as Tevatron and LHC.
Relevant results have been obtained by the CDF and the D0
collaborations~\cite{Aaltonen:2010jj,Abazov:2007ah} and more recently by the ATLAS
and the CMS collaborations~
\cite{Aad:2011yg,Khachatryan:2010fa,Chatrchyan:2011dx,Aad:2011fe} in terms of the
$W_R$ or $W^{\prime}$ bosons or heavy charged right-handed bosons searches (not
necessarily LRSM).
So far, most of the works are done in the Sequential Standard Model (SSM) in which
the coupling of $W^\prime WZ$ is set to be zero, and the $W^\prime$ has the same couplings to fermions as the SM $W$ boson\cite{Altarelli:1989ff}.
In the LRSM, the coupling of $W_2$ (or say $W^\prime$) to $W_1Z_1$ (or say $WZ$) is zero, because the left- and right-handed quantum numbers of $Z_1$ is zero, then it can not couple to $W_LW_R$ and hence not to $W_1W_2$. In addition, if $g_R=g_L$, the coupling of $W_2$ to fermions is identical to the one of $W_1$ to fermions. Therefore,the LRSM in the case of $g_R/g_L=1$ corresponds to the SSM.
The current limit for $M_{W^\prime}$ in the case of the SSM is obtained as  $M_{W^\prime}>2.5$ TeV  at the $95\%$ confidence level  by ATLAS and CMS,  using its decay into a charged lepton and a neutrino at $\sqrt{s}=7$ TeV~\cite{newlhcconstraints1,newlhcconstraints2}.

\begin{figure}[thpb]
\begin{center}
\includegraphics[scale=0.4]{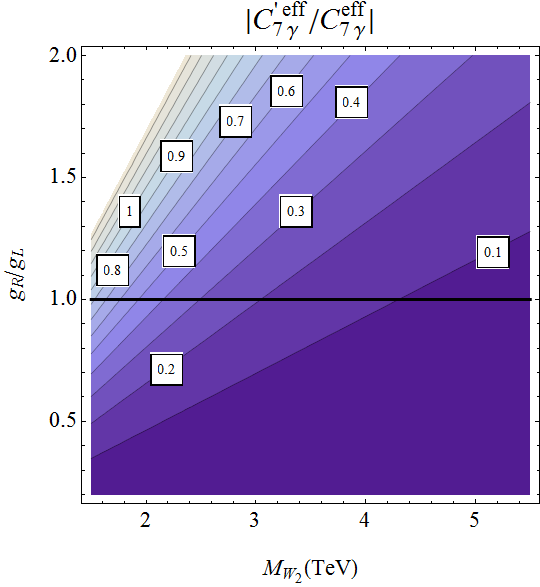}
\caption{ The dependence of $|C_7^\prime/C_7|$ on the mass $M_{W_2}$ and the gauge coupling ratio $g_R/g_L$. }\label{fig:gRgL}
\end{center}
\end{figure}

\begin{figure}[thpb]
\begin{center}
\includegraphics[scale=0.4]{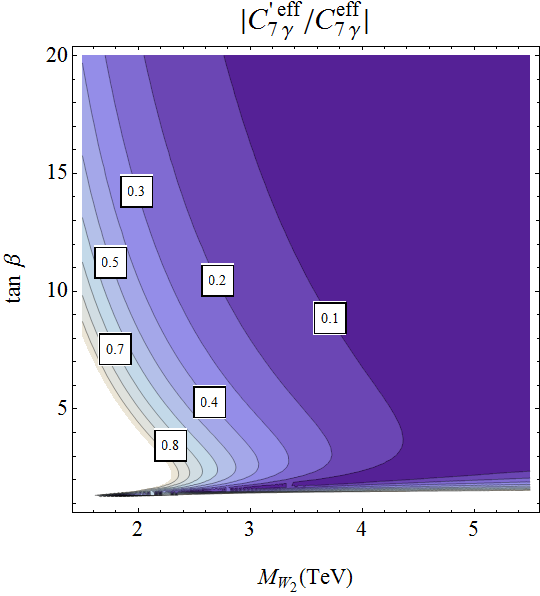}
\caption{ The dependence of $|C_7^\prime/C_7|$ on  $M_{W_2}$ and $\tan\beta$. }\label{fig:tanbetaM2}
\end{center}
\end{figure}

Now, let us include this limit to our analysis in the previous sections.
In Figure~\ref{fig:gRgL}, we plot the dependence of $|C_7^\prime/C_7|$ on the mass $M_{W_2}$ and on the gauge coupling ratio $g_R/g_L$, i.e.  Eq.~(\ref{c7poc7num}).
In this figure, we fix $\tan\beta=10$ and $\sin\theta_{23}=1$ (which is allowed by the  $\Delta F=2$ observables).
Looking at the horizontal line, $g_R/g_L=1$, we find that the LHC limit $M_W>2.5$ TeV allows  $|C_7^\prime/C_7|$ to be $\lesssim 0.3$, as we have seen in the previous section.
As the LHC experiments will push this mass limit to a higher value, the possible size for $|C_7^\prime/C_7|$ will decrease rapidly.
Let's say, for $M_W>5$ TeV, $|C_7^\prime/C_7|\lesssim 0.1$. On the other hand, in order to investigate the full range of the LRSM parameter space, we also need the mass limit for different values of $g_R/g_L$.
For example, if we assume a larger value of $g_R/g_L$, a limit on $|C_7^\prime/C_7|$ can be relaxed.
In this sense, these two experiments are complementary.
We should also note that so far, we have assumed $\tan\beta=10$ though
$|C_7^\prime/C_7|$ strongly depends on $\tan\beta$  (see Figure~\ref{fig:tanbetaM2}). From Figure~\ref{fig:tanbetaM2}, the odd behavior at $\tan\beta\sim1$ is from the charged Higgs contribution to $|C_7^\prime/C_7|$, which is divergent at $\tan\beta=1$.
Thus, measurements of $|C_7^\prime/C_7|$ and the mass limit from the direct search should be used exhaustively  to constrain the whole parameter space.


\section{Conclusions}

In this work, we investigate  possible right-handed photon polarization in the $b\to s\gamma$ processes with the allowed range of parameters in the Left-Right Symmetric Model (LRSM).
The circular-polarization of the photon in the radiative $B$ decays
are dominantly left-handed in the Standard Model (SM).
But we find that in the LRSM, the right-handed polarization can be enhanced by the  $W_L-W_R$ mixing contributions because of the helicity flip on the internal top quark in the penguin diagrams and the enhancement by the CKM factor $V_{ts}^R/V_{ts}^L$.
Note that for the right-handed CKM matrix, we extended the previously proposed parameterization  in the literature by incorporating the additional CP violating phase.

We first  constrained the parameter space of the LRSM by the observables of  $b\to s\ga$ and $\Delta F=2$ processes, such as the inclusive branching ratio of $\overline B\to X_s\gamma$, the mass differences and CP asymmetries in the $K^0-\overline K^0$, $B_d^0-\overline B_d^0$ and $B_s^0-\overline B_s^0$ mixing systems. In order to find the allowed parameter space, we consider not only the experimental errors, but also the theoretical uncertainties carefully.
We find that the observables in the neutral meson mixing systems are significantly affected by the tree-level contributions from the heavy neutral Higgs bosons $H_1^0, A_1^0$, as well as the box diagrams with the charged gauge bosons $W_R^\pm$.

Our numerical analysis shows that the ratio of right-handed and left-handed polarization of the photon in the $b\to s\ga$ processes, i.e. $|C_{7\ga}^\prime/C_{7\ga}|$, could be much larger than the SM prediction, with its value as large as (0.5, 0.35, 0.25) for the mass of $W_2^\pm$ as (2.0, 2.5, 3.0) TeV in the case of $g_R/g_L=1$ and $\tan\beta=10$.

We also investigate the impact of the future photon polarization measurements on the LRSM, such as
the time-dependent CP asymmetry in the decay of $\overline B\to K_S \pi^0\ga$ ($S_{K_S\pi\ga}$), the angular distribution in the $\overline B\to \overline K_1\ga\to K\pi\pi\ga$ decay ($\lambda_\ga$), and the angular analysis of $\overline B\to \overline K^*(\to K\pi)\ell^+\ell^-$  ($A_T^{(2)}, A_T^{(im)}$). We find that the LRSM can lead to large enough values of $S_{K_S\pi\ga}$, $\lambda_\ga$, $A_T^{(2)}$ and  $A_T^{(im)}$ which are easily accessible by the LHCb and the future BELLE II experiments. We finally showed  that the measurement of the $b\to s\gamma$ photon polarization plays a complementary role to  the direct search of the right-handed charged gauge boson at the LHC.

\section*{Acknowledgements}
The  work of E.K. and F.S.Y. was supported in part by the ANR contract ``LFV-CPV-LHC'' ANR-NT09-508531 and France-China corporation of IN2P3/ CNRS FCPPL-LIA. The work of C.D.L. and F.S.Y. is partly supported by the National Science Foundation of China under the Grant No. 11228512, 11235005 and 11311130102.

\appendix

\section{Interactions in the LRSM}\label{app:Lagrangian}

The charged-current associated with $W_{1,2}^\pm$ bosons is
\be
\mathcal{L}_{cc}^W&=&\bar u_i\ga^\mu\left({g_L\over\sqrt 2}V_{ij}^L\cos\zeta P_L+{g_R\over\sqrt 2}V_{ij}^R\sin\zeta e^{i w}P_R\right)d_j W_{1\mu}^+
\non\\
&&+\bar u_i\ga^\mu\left(-{g_L\over\sqrt 2}V_{ij}^L\sin\zeta e^{-i w} P_L+{g_R\over\sqrt 2}V_{ij}^R\cos\zeta P_R\right)d_j W_{2\mu}^++h.c..
\ed
where $\zeta$ is the mixing angle between $W_L^\pm$ and $W_R^\pm$, $P_{L,R}\equiv \frac{1\mp\gamma_5}{2}$, $g_{L,R}$ are the gauge coupling constants in the $SU(2)_L$ and $SU(2)_R$ groups, $V_{ij}^{L,R}$ are the left- and right-handed CKM matrices, and $w$ is the spontaneous CP violating phase.
The charged-current associated with charged Goldstone bosons is
\be
\mathcal{L}_{cc}^G&=&{1\over\sqrt{2}M_{W_1}}\bar u_i\bigg[\big(m_{u_i}{g_L}V_{ij}^L\cos\zeta -m_{d_j}{g_R}V_{ij}^R\sin\zeta e^{i w}\big)P_L
\non\\&&-\big(m_{d_j}g_LV_{ij}^L\cos\zeta-m_{u_i}g_R V_{ij}^R\sin\zeta e^{i w}\big)P_R\bigg]d_j G_{1}^+
\non\\
&+&{1\over\sqrt{2}M_{W_2}}\bar u_i\bigg[\big(m_{d_j}{g_L}V_{ij}^L\sin\zeta e^{-i w} +m_{u_i}{g_R}V_{ij}^R\cos\zeta \big)P_R
\non\\&&-\big(m_{u_i}g_LV_{ij}^L\sin\zeta e^{-i w}+m_{d_j}g_R V_{ij}^R\cos\zeta\big)P_L\bigg]d_j G_{2}^+
\non\\&+&h.c.,
\ed

The charged-current with charged Higgs boson is
\be
\mathcal{L}_{cc}^H&=&N_{H^+}\bar u_i\bigg[\big({\tan2\beta}e^{-i w}m_{d_j}V_{ij}^L -{1\over\cos2\beta}m_{u_i}V_{ij}^R\big)P_R
\non\\&&-\big({\tan2\beta}e^{-i w}m_{u_i}V_{ij}^L -{1\over\cos2\beta}m_{d_j}V_{ij}^R\big)P_L\bigg]d_j H^+
\non\\&&+h.c.
\ed
with $N_{H^+}={1\over v}(1-{1\over4}\epsilon^2\cos^22\beta)$, where $v$, $\epsilon$, and $\beta$ are defined in Eqs. (\ref{eq:v}) and (\ref{eq:tanbeta}).

{
The Lagrangian of couplings between charged gauge bosons, charged Goldstone bosons and photons are
\be
\mathcal{L}_{WWA}&=&ie\Big[A_\mu W_{1\nu}^+\left(\partial_\mu W_{1\nu}^--\partial_\nu W_{1\mu}^-\right)+A_\mu W_{1\nu}^-\left(\partial_\nu W_{1\mu}^+-\partial_\mu W_{1\nu}^+\right)
\non\\&&+W_{1\mu}^+ W_{1\nu}^-\left(\partial_\mu A_{\nu}-\partial_\nu A_{\mu}\right)\Big]
+(W_1\leftrightarrow W_2),
\ed
\be
\mathcal{L}_{GGA,HHA}=ieA_\mu\Big[\left(G_1^-\partial_\mu G_1^+-G_1\partial_\mu G_1^-\right)+(G_1^\pm\leftrightarrow G_2^\pm)+(H^\pm\leftrightarrow H^\pm)\Big],
\ed
\be
\mathcal{L}_{GWA}=eM_{W_1}A_\mu\left(G_1^+W_{1\mu}^-+G_1^-W_{1\mu}^+\right)
+eM_{W_2}A_\mu\left(G_2^+W_{2\mu}^-+G_2^-W_{2\mu}^+\right).
\ed
}

The Yukawa coupling between heavy neutral Higgs and quarks are \cite{Zhang:2007da}
\be
\mathcal{L}_{\rm NH}&=&(\sqrt{2}G_F)^{1/2}\bigg[\bar u_{Li} \left( V_L M_D V_R^\dag\right)_{ij}\left(H^0-i A^0\right)u_{Rj}+\bar d_{Li} \left( V_L^\dag M_U V_R\right)_{ij}\left(H^0+i A^0\right)d_{Rj}\bigg]\non\\
&&+h.c.
\ed

The effective Hamiltonian of $b\to c\bar cs$ in the LRSM at the tree level is
\be\label{appA:tree}
\mathcal{H}_{\rm tree}&=&{g_L^2\over2}V_{cs}^{L*}V_{cb}^L{1\over
 M_1^2}\bar c\ga^\mu P_L b\cdot\bar s
 \ga_\mu P_L c
 +{g_R^2\over2}V_{cs}^{R*}V_{cb}^R{1\over
 M_2^2}\bar c\ga^\mu P_R b\cdot\bar s
 \ga_\mu P_R c
 \non\\
 &&+{g_Lg_R\over2}V_{cs}^{L*}V_{cb}^Re^{i\omega}{1\over
 M_1^2}\sin\zeta\bar c\ga^\mu P_L b\cdot\bar s
 \ga_\mu P_R c
 \non\\
 &&+{g_Lg_R\over2}V_{cb}^L(V_{cs}^{R}e^{i\omega})^*{1\over
 M_1^2}\sin\zeta\bar c\ga^\mu P_R b\cdot\bar s
 \ga_\mu P_L c.
\ed


\section{Wilson coefficients $C_{7\gamma}$ and $C_{7\gamma}^{\prime}$}\label{app:C7}
In the LRSM, the Wilson coefficients for the $b\to s\ga$ processes are
\be
C_{7\gamma}(\mu_R)={1\over2}&\Bigg[&\cos^2\zeta A_{\rm SM}(x_t)+\sin^2\zeta {M_1^2\over M_2^2}A_{\rm SM}(\tilde x_t)\non\\&+&{m_t\over m_b}{g_R\over g_L}{V_{tb}^{R}\over V_{tb}^L}\sin\zeta\cos\zeta e^{i w}\Big(A_{\rm LR}(x_t)-{M_1^2\over M_2^2}A_{\rm LR}(\tilde x_t)\Big)\non\\
&+&{ {m_c\over m_b}{g_R\over g_L}{V_{cs}^{L*}V_{cb}^{R}\over V_{ts}^{L*}V_{tb}^L}\sin\zeta\cos\zeta e^{i w}\Big(A_{\rm LR}(x_c)-{M_1^2\over M_2^2}A_{\rm LR}(\tilde x_c)\Big)}\non\\
&+&{m_t\over m_b}{\tan2\beta\over\cos2\beta}e^{i w}{V_{tb}^R\over V_{tb}^L}A_H^1(y)+\tan^22\beta A_H^2(y)\Bigg]\label{eq:c7lall}
\\
C_{7\gamma}^\prime(\mu_R)={1\over2}&\Bigg[&{g_R^2\over g_L^2}{V_{ts}^{R*}V_{tb}^R\over
V_{ts}^{L*}V_{tb}^L}\Big(\sin^2\zeta A_{\rm SM}(x_t)+\cos^2\zeta {M_1^2\over M_2^2}A_{\rm SM}(\tilde x_t)\Big)
\non\\&+&{m_t\over m_b}{g_R\over g_L}{V_{ts}^{R*}\over V_{ts}^{L*}}\sin\zeta\cos\zeta e^{-i w}\Big(A_{\rm LR}(x_t)-{M_1^2\over M_2^2}A_{\rm LR}(\tilde x_t)\Big)\non\\\label{eq:c7rall}
&+&{ {m_c\over m_b}{g_R\over g_L}{V_{cs}^{R*}V_{cb}^{L}\over V_{ts}^{L*}V_{tb}^L}\sin\zeta\cos\zeta e^{-i w}\Big(A_{\rm LR}(x_c)-{M_1^2\over M_2^2}A_{\rm LR}(\tilde x_c)\Big)}\non\\
&+&{m_t\over m_b}{\tan2\beta\over\cos2\beta}e^{-i w}{V_{ts}^{R*}\over V_{ts}^{L*}}A_H^1(y)+{V_{ts}^{R*}V_{tb}^R\over
V_{ts}^{L*}V_{tb}^L}{1\over\cos^22\beta} A_H^2(y)\Bigg]\label{eq:c7r}
\ed
where $x_i={m_i^2/ M_{W_1}^2}$, ${\tilde x_i=m_i^2/M_{W_2}^2}$, $i=c,t$, $y={m_t^2\over M_{H}^2}$, with the loop functions
\be
A_{\rm SM}(x)&=&{-8x^3-5x^2+7x\over12(x-1)^3}+{3x^3-2x^2\over2(x-1)^4}\ln x=2C_{2\gamma}^{\rm SM},
\\
A_{\rm LR}(x)&=&{-5x^2+31x-20\over6(x-1)^2}-{3x^2-2x\over(x-1)^3}\ln x,
\\
A_{H}^1(y)&=&{-5y^2+3y\over6(y-1)^3}+{3y^2-2y\over3(y-1)^2}\ln y,
\\
A_{H}^2(y)&=&{22y^3-53y^2+25y\over36(y-1)^3}-{3y^3-8y^2+4y\over6(y-1)^4}\ln y.
\ed

{  Note that there is a constant term in $A_{\rm LR}(x)$, that is $A_{\rm LR}(x\to0)=-10/3$, because there is no GIM mechanism for the corresponding terms in Eqs.(\ref{eq:c7lall},\ref{eq:c7rall}) with one left-handed CKM matrix element and one right-handed CKM matrix element. Whilst the constant term in $A_{\rm SM}(x)$ is canceled by the GIM mechanism. Therefore we have to consider the contribution from charm quarks in the loops, as shown in Eqs.(\ref{eq:c7lall},\ref{eq:c7rall}). The up quark contribution is neglected due to its tiny mass. Numerically,  in Eq.(\ref{eq:c7lall}), $m_t V_{ts}^L A_{\rm LR}(x_1^t)=10.5$, while $m_c V_{cs}^L A_{\rm LR}(x_1^c)=-4.2$, it is clear that the charm contribution has to be considered.} We could neglect the charm contribution in Eq.(\ref{eq:c7rall}) because of a suppression by $m_c V_{cb}^L$.

\section{Box loop functions for LRSM}\label{app:box}
The box loop functions for the LRSM in Eqs.(\ref{Kbox}) are the following:
 \be\label{Kboxinte}
S_{LL}(x_i,x_j)&=&F(x_i,x_j)+F(x_u,x_u)-F(x_i,x_u)-F(x_j,x_u),
\\
S_{LR}(x_i,x_j,\eta)&=&8 r^2\eta\sqrt{x_i x_j}\left[(4+ x_i x_j\eta)I_1(x_i,x_j,\eta)-(1+\eta)I_2(x_i,x_j,\eta)\right],
\\
S_{RR}(\tilde x_i,\tilde x_j)&=&  r^4\eta S_{LL}(\tilde x_i,\tilde x_j),
\ed
with $r\equiv g_R/g_L$, $\eta\equiv M_{W_1}^2/M_{W_2}^2$, $x_i\equiv m_i^2/M_{W_1}^2$, $\tilde x_i\equiv m_i^2/M_{W_2}^2$, and $i=u,c,t$,
\be
F(x_i,x_j)=(4+x_ix_j)I_2(x_i,x_j,1)-8x_ix_jI_1(x_i,x_j,1),
\ed
\be
I_1(x_i,x_j,\eta)&=&{x_i\ln(x_i)\over(1-x_i)(1-x_i\eta)(x_i-x_j)}+(i\leftrightarrow j)-{\eta\ln(\eta)\over(1-\eta)(1-x_i\eta)(1-x_j\eta)},
\\
I_2(x_i,x_j,\eta)&=&{x_i^2\ln(x_i)\over(1-x_i)(1-x_i\eta)(x_i-x_j)}+(i\leftrightarrow j)-{\ln(\eta)\over(1-\eta)(1-x_i\eta)(1-x_j\eta)}.
\ed

\section{Bag parameters}\label{app:bag}
The bag parameters are always calculated in the regularization-independent (RI) scheme by Lattice QCD.
But the Wilson coefficients are always obtained in the naive dimension regularization (NDR) scheme in the perturbative theory.
Therefore the bag paraemters obtained by the lattice QCD should be transformed into the NDR scheme.
In this work, we use the bag parameters in the NDR scheme for Kaon mixing at $\mu=2$GeV given in \cite{Bertone:2012cu}
 \be
B_1^{\rm VLL}(K)_{\rm NDR}=0.52,&&
\nonumber\\
 B_1^{\rm LR}(K)_{\rm NDR}=0.63,&&B_2^{\rm LR}(K)_{\rm NDR}=0.82,
\nonumber\\
B_1^{\rm SLL}(K)_{\rm NDR}=0.54,&&B_2^{\rm SLL}(K)_{\rm NDR}=0.27,
\ed
and for the $B_d$ and $B_s$ case at $\mu_b=4.6$GeV  given in \cite{Becirevic:2001xt}
 \be
B_1^{\rm VLL}(B)_{\rm NDR}=0.87,&&
\nonumber\\
 B_1^{\rm LR}(B)_{\rm NDR}=1.72,&&B_2^{\rm LR}(B)_{\rm NDR}=1.15,
\nonumber\\
B_1^{\rm SLL}(B)_{\rm NDR}=0.84,&&B_2^{\rm SLL}(B)_{\rm NDR}=0.79,
\ed
with an approximation that the bag parameters are the same for $B_d$ and $
B_s$ systems within the uncertainties.

\section{Procedure to find the allowed parameter space}\label{app:fit}
To simplify the task of finding the allowed parameter space, we introduce the following $\chi^2$ including  $|\epsilon_K|$, $\Delta M_s$, $\phi_s$, $Br(B\to X_s\gamma)$ as well as the variables listed in Table~\ref{Tab:1}:
 \be\label{eq:chi2}
 \chi^2&=&{(|\epsilon_K^{\rm th}|-|\epsilon_K^{\rm exp}|)^2\over (\sigma_{|\epsilon_K|}^{\rm exp})^2}+{(\Delta M_s^{\rm th}-\Delta M_s^{\rm exp})^2\over (\sigma_{\Delta M_s}^{\rm exp})^2}+{(\phi_s^{\rm th}-\phi_s^{\rm exp})^2\over (\sigma_{\phi_s}^{\rm exp})^2}+{(Br_{b\to s\ga}^{\rm th}-Br_{b\to s\ga}^{\rm exp})^2\over (\sigma_{Br_{b\to s\ga}}^{\rm exp})^2}\non\\
 &+&\sum_{i}{(var_i-var_i^{\rm exp})^2\over (\sigma_{var_i})^2},
 \ed
where the $var_i$ are $|V_{td}|$, $|V_{ts}|$, $|V_{tb}|$,  $\beta_{\rm CKM}$, $B_{2}^{LR}(K)$, $B_1^{VLL}(B)$, and $B_2^{LR}(B)$. In addition, as our target is to maximize $|C_{7\gamma}^{\prime}/C_{7\gamma}|$, which can be achieved only when $\theta_{23}\simeq \pm \pi/2$, we also include $\theta_{23}$ to be a variable and vary in the region of $\theta_{23}=-{\pi\over2}\pm0.3$ (we found that the result does not change for $\pm \pi/2$).
For the given point on the $(\phi_1, \phi_2)$ space, which is chosen in $[-\pi, \pi]$ with $\pi/50$ interval, we minimize this $\chi^2$. We chose the point which give the $\chi^2$ to be less than a certain value as an allowed point. This value is chosen conservatively so that any components of $\chi^2$ in Eq.~(\ref{eq:chi2}) do not exceed its three sigma range.

\section{Observables to determine the photon polarization}\label{appG}
We describe the three observables proposed in order to determine the photon polarization.
\begin{description}
\item[i)] {The time dependent $CP$-asymmetry, $S_{K_S}\pi^0\ga$}: \\
{The time-dependent $CP$-asymmetry in the radiative decays of the neutral $B$ mesons is an indirect method to measure the photon polarization. It arises from the interference between $\overline B\to f_{CP}\ga$ and $\overline B\to B\to f_{CP}\ga$ amplitudes where $f_{CP}$ is a CP eigenstate. The amplitudes with the photon with left- and right-handed helicity can not interfere with each other since the photon polarization is measurable in principle. Then the $CP$-asymmetry is defined as \cite{kpiga1,Atwood:2004jj}
\be
A_{CP}(t)\equiv {\Gamma(\overline B(t)\to f_{CP}\ga)-\Gamma(B(t)\to f_{CP}\ga) \over \Gamma(\overline B(t)\to f_{CP}\ga)+\Gamma(B(t)\to f_{CP}\ga)}\approx S_{f_{CP}\ga}\sin(\Delta mt),
\ed
with
\be
S_{f_{CP}\ga}\equiv \xi {2Im\left[e^{-i\phi_M}\mathcal{M}(\overline B\to f_{CP}\ga_L)\mathcal{M}(\overline B\to f_{CP}\ga_R)\right]\over \left|\mathcal{M}(\overline B\to f_{CP}\ga_L)\right|^2+\left|\mathcal{M}(\overline B\to f_{CP}\ga_R)\right|^2}\approx \xi {2Im\left[e^{-i\phi_M} C_{7\ga}C_{7\ga}^\prime\right]\over |C_{7\ga}|^2+|C_{7\ga}^\prime|^2},
\ed
where $\xi(=\pm1)$ is the $CP$-eigenvalue of $f_{CP}$, and $\phi_M$ is the phase in the $B-\overline B$ mixing, which in the SM is $\phi_d=2\beta_{\rm CKM}\simeq0.75$ and $\phi_s\simeq0$ for the $B_d^0$ and $B_s^0$ mixing respectively. In the SM $S_{f_{CP}\ga}$ are always quite small because of the smallness of $C_{7\ga}^{\prime{\rm SM}}$. Specifically, the SM prediction for the $\overline B^0\to \overline K^{(*0)}(\to K_S\pi^0)\ga$ decay is \cite{Ball:2006eu}
\be
S_{K_S\pi^0\ga}^{\rm SM}=-(2.3\pm1.6)\%,
\ed
while its current world average value is \cite{HFAG}
\be
S_{K_S\pi^0\ga}^{\rm exp}=-0.16\pm0.23.
\ed
In the future Belle II experiment, the error of $S_{K_S\pi^0\ga}$ will be significantly reduced down to 2$\%$.

\item[ii)] The polarization parameter $\lambda_\gamma$ using the $\overline B\to \overline K_1\ga\to K\pi\pi\ga$ process: \\
Compared to the $CP$-asymmetry, a direct method to measure the photon polarization is to study the three-body decays of $K_1$ ($\overline B\to \overline K_1\ga\to K\pi\pi\ga$) \cite{k1gamma1,k1gamma2}. The polarization can be studied by the angular distribution in the $\overline B\to \overline K_1\ga\to K\pi\pi\ga$ decay, extracting the polarization parameter $\lambda_\ga$
\be
\lambda_\ga\equiv {\left|\mathcal{M}(\overline B\to \overline K_{1R}\ga_R)\right|^2-\left|\mathcal{M}(\overline B\to \overline K_{1L}\ga_L)\right|^2 \over \left|\mathcal{M}(\overline B\to \overline K_{1}\ga)\right|^2}\approx {\left|C_{7\ga}^\prime\right|^2-\left|C_{7\ga}\right|^2 \over \left|C_{7\ga}^\prime\right|^2+\left|C_{7\ga}\right|^2}.
\ed
In the SM, $\lambda_\ga\simeq-1(+1)$ for $\overline B(B)$ decays. In the future, Belle II / LHCb experiment can measure this observable at the precision of $\sigma(\lambda_\gamma)\simeq0.2$.

\item[iii)]
The angular analysis of $\overline B\to \overline K^*(\to K\pi)\ell^+\ell^-$: \\
Among various angular coefficients, the most interesting observables are the transverse asymmetries defined as \cite{Kruger:2005ep}\cite{Becirevic:2011bp}
\be
A_T^{(2)}(q^2)={I_3(q^2)\over2I_2^s(q^2)}, ~~~~~A_T^{(im)}(q^2)={I_9(q^2)\over2I_2^s(q^2)},
\ed
where $q^2$ is the $\ell^+\ell^-$ invariant mass, $I_i(q^2)$ are differential decay amplitudes which can be found in \cite{theoryworks1}\cite{Becirevic:2012dx}. In the limit of $q^2=0$, only the $O_{7\ga}^{(\prime)}$ operators contribute to this process, so that the above asymmetries are only related to $C_{7\ga}^{(\prime)}$\cite{Becirevic:2012dx},
\be
A_T^{(2)}(0)&=&{2Re\left[C_{7\ga}C_{7\ga}^{\prime*}\right]\over \left|C_{7\ga}\right|^2+\left|C_{7\ga}^\prime\right|^2},\\
A_T^{(im)}(0)&=&{2Im\left[C_{7\ga}C_{7\ga}^{\prime*}\right]\over \left|C_{7\ga}\right|^2+\left|C_{7\ga}^\prime\right|^2}.
\ed
}
In the future, Belle II / LHCb experiments can measure this observable at the precision of $\sigma(A_T^{(2)})\simeq\sigma(A_T^{(im)})\simeq0.2$.
\end{description}

\end{document}